\documentclass{article}
\usepackage{graphicx} 
\usepackage[margin=1.25in]{geometry}
\usepackage{amsmath}
\usepackage{amssymb} 
\usepackage{xcolor}
\usepackage[square,numbers,sort&compress]{natbib}
\usepackage{titling}
\usepackage{fontawesome5}
\definecolor{linkgreen}{RGB}{34,102,58}
\usepackage[colorlinks=true,linkcolor=linkgreen,citecolor=linkgreen,urlcolor=linkgreen]{hyperref}

\def\etab{\boldsymbol{\eta}}

\def\gb{\boldsymbol{g}}

\def\xb{\boldsymbol{x}}

\def\zb{\boldsymbol{z}}

\def\d{\mathrm{d}}

\def\RR{\mathbb{R}}  
\def\EE{\mathbb{E}}

\newcommand{\diff}[3]{\frac{\partial^{#3}#1}{\partial #2^{#3}}}

\title{Scaling Transferable Coarse-graining with Mean Force Matching}
\author{Abigail Park$^\dagger$, Shriram Chennakesavalu$^\dagger$, Grant M. Rotskoff$^{\dagger,\ddagger}$\thanks{Corresponding author: \href{mailto:rotskoff@stanford.edu}{rotskoff@stanford.edu}} \\
$\dagger$ Department of Chemistry, Stanford University, Stanford, CA, 94305 \\
$\ddagger$ ICME, Stanford University, Stanford, CA, 94305}

\begin{document}
\maketitle

\begin{abstract}
    Coarse-grained molecular dynamics often sacrifices accuracy and transferability for computational efficiency, but the use of machine learned potentials is helping coarse-grained models attain performance on par with atomistic molecular dynamics.
    Nevertheless, developing representations of the coarse-grained potential energy surface faces severe scaling challenges due to the extreme data demands of widely used ``bottom-up'' coarse-graining objectives. 
    In this work, we show that mean force matching, a strategy for training thermodynamically consistent coarse-grained models, requires 50$\times$ fewer training samples and 87\% less total atomistic simulation time, while obtaining better accuracy on the potential of mean force for unseen proteins compared to other commonly used objectives.
    By systematically removing noise from the objective function, we demonstrate that it is possible to scale machine learning architectures for coarse-graining, enabling highly accurate and transferable models.
    We show the advantages of mean force matching both theoretically and through exhaustive benchmarking using thermodynamic consistency as the primary metric of accuracy. 
\end{abstract}

\begin{center}
\small \faGithub\ \href{https://github.com/rotskoff-group/transferable-cg}{GitHub} \quad|\quad \faDatabase\ \href{https://huggingface.co/datasets/abpark/transferable-cg}{Hugging Face}
\end{center}


\section{Introduction}
Coarse-grained models of proteins are widely used to study complex biomolecular phenomena that are infeasible to simulate with atomistic molecular dynamics (MD), but these models often involve a trade-off between transferability across systems and accuracy~\cite{wagner_representability_2016}.
We typically design coarse-grained models to recapitulate a set of properties of interest, either optimizing alignment with experimental observables (top-down CG) or an underlying physics-based model (bottom-up CG)~\cite{pak_advances_2018,jin_bottom-up_2022}.
Models within these two categories have proven useful for studying diverse sets of biomolecular systems, including protein folding~\cite{clementi_coarse-grained_2008}, biomolecular condensates of intrinsically disordered proteins~\cite{tesei_accurate_2021,tesei_conformational_2024}, cytoskeletal networks~\cite{mccullagh_unraveling_2014}, and mesoscale self-assembly~\cite{grime_coarse-grained_2016}.
However, many of these ``bottom-up'' models are trained extensively on system-specific atomistic MD, which creates a significant barrier to deployment. 
Only recently have transferable CG models demonstrated robust qualitative agreement with thermodynamic properties as measured in MD~\cite{majewski_machine_2023,charron_navigating_2025}, but best practices for training and scaling these models remain unclear. 
 
Machine learning has played an important role in improving transferability in CG potentials, but the desiderata in both model architecture and data set size have not been clearly evaluated due to the high computational cost of training CG models.
Thermodynamic consistency~\cite{izvekov_multiscale_2005, noid_multiscale_2008}---that is, capturing the potential of mean force for the coarse-grained degrees of freedom---is the primary objective of physics-based coarse-graining, and it can, in principle, be achieved using a variety of distinct objectives~\cite{shell_relative_2008}.
Force matching, which uses noisy, instantaneous projected forces from atomistic MD has proved a popular choice for building thermodynamically consistent CG models, but requires large amounts of correlated data to mitigate the noise~\cite{durumeric_adversarial-residual-coarse-graining_2019,kramer_statistically_2023} or post-processing and data augmentation (which can further increase training costs).
Alternative objectives based on the widely used generative modeling loss functions for denoising diffusion probabilistic models~\cite{song_score-based_2022,hyvarinen_estimation_2005} have been combined with force matching to amplify the signal in the data~\cite{arts_two_2023}, but the relative performance of these approaches is not clear.
To more comprehensively benchmark model design choices alongside training objectives, it is imperative that we build more computationally tractable strategies for parameterizing CG models. 

The high training costs of CG models create bottlenecks to scaling the parameter count and complexity of model architectures and data set size, both of which have been enormously productive for machine learned interatomic potentials (MLIPs) that aim to learn potential energy surfaces from electronic structure calculations~\cite{deringer_machine_2019,wood_uma_2025}. 
As is well-documented across many domains in the machine learning literature~\cite{bahri_explaining_2024}, increases in model size and training data lead to predictable performance improvements and enable compute-optimal resource allocation~\cite{hoffmann_training_2022} due to the strong adherence to neural scaling laws.
While we anticipate that highly scalable MLIP architectures should lead to improved transferability and accuracy, at present, the poor scalability of both force matching and score matching objectives presents an intractable computational barrier to evaluating similar trends for CG model training.

In this paper, we describe an approach that massively improves scalability of training transferable coarse-grained models, which we demonstrate both theoretically and through extensive benchmarking.
The strategy we propose is extremely simple: we reduce the variance of the mean forces by averaging over constrained MD simulations.
To achieve sufficient sample diversity, this approach relies on generating a diverse set of initial atomistic structures from simulations at a variety of temperatures or enhanced sampling, but we show that the total computational cost of data generation and training is markedly lower than force matching.
A straightforward mathematical analysis shows that mean force matching provably diminishes the variance and leads to greater training signal than instantaneous force matching or score matching.

By carefully constructing data sets and training protocols, we compare three distinct loss functions (force matching, score matching, and mean force matching) in the context of training transferable coarse-grained models.
The benchmark we establish also allows us to compare training costs, inference costs, and the zero-shot accuracy of free energy surfaces computed with models trained using these three objectives and distinct neural network architectures.
Throughout, we find that mean force matching achieves the highest accuracy on unseen proteins, despite requiring only a small fraction of the total training data, even when accounting for the cost of generating initial structures.
Our approach sets the stage for scaling both data and CG models to further improve accuracy.

Our results indicate that message-passing MLIPs such as MACE and eSEN achieve the best performance, though the different architectures we consider have highly variable ``inference-time'' computational costs.
Efficient implementations of MACE using NVIDIA's cuEquivariance~\cite{noauthor_cuequivariance_2025} package, in particular, strike the best balance between error and efficiency.
That said, the structure of the MACE architecture limits scalability in depth because each additional layer corresponds to an increase in body-order of a many-body expansion~\cite{batatia_mace_2023}.
While we clearly observe that increasing model size leads to predictable accuracy improvements, current MLIPs remain too computationally expensive for transformative acceleration of sampling relative to atomistic MD. 

The models that we build accurately recapitulate the folded and unfolded ensembles of a variety of proteins with distinct fold topologies ``zero-shot'', i.e., never having seen any system specific data.
Furthermore, our CG models yield results in good agreement with atomistic MD for proteins with low sequence homology to any proteins in the training set. 
While the free energy surfaces that we compute capture the thermodynamically relevant states, they lack quantitative agreement, especially near statistically rare transition states.
However, combining these coarse-grained models with statistically rigorous back-mapping strategies offers a route to exactly correct this discrepancy \cite{chennakesavalu_ensuring_2023}.
Viewed more broadly within the context of generative models for sampling, such as Boltzmann generators, we believe physics-based coarse-graining offers a powerful strategy to explore a ``latent space'' that will not mode collapse on system specific training data. 

\begin{figure}
    \centering
    \includegraphics[width=\linewidth]{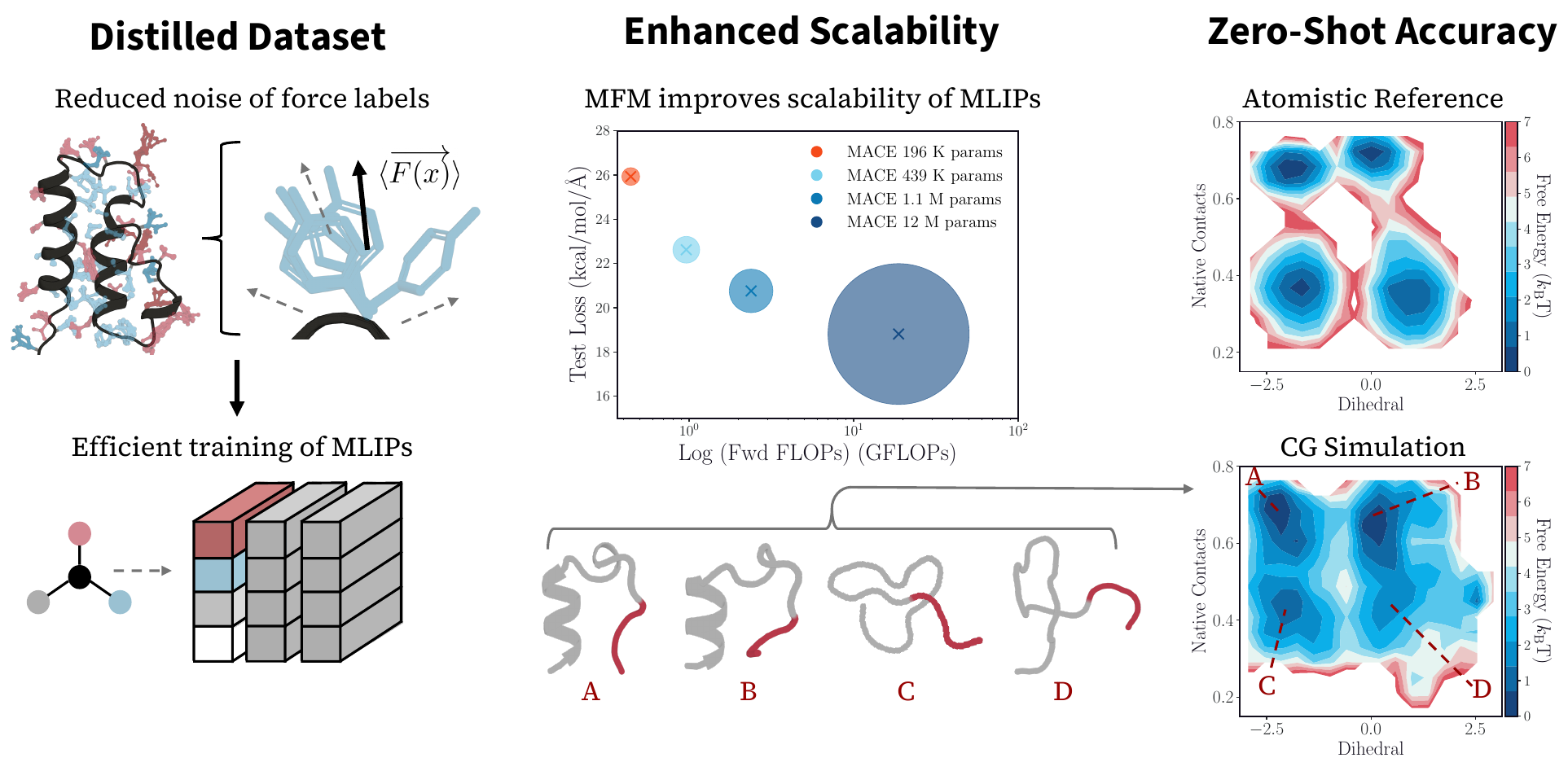}
    \caption{Mean Force Matching (MFM) enables efficient training of transferable coarse-grained protein force fields with enhanced scalability and zero-shot accuracy. (Left) Distilled Dataset: mean forces extracted from constrained atomistic MD simulations reduce noise compared to instantaneous forces, enabling efficient MLIP training.  (Center) Enhanced Scalability: test loss scaling of MFM as a function of computational budget for various model sizes. (Right) Zero-Shot Accuracy: MFM CG model produces free energy surface that closely resembles atomistic reference for Trp-cage}
    \label{fig:placeholder}
\end{figure}

\section{Thermodynamically Consistent CG and Mean Force Matching}

Consider a system with coordinates $\xb\in \RR^{n}$. We seek a low-dimensional, or coarse-grained, representation of this system in some possibly nonlinear subspace, which we represent as $\zb\in \RR^{k}$ with $k\ll n$.
We assume that the probability distribution of the fine-grained system is given by,
\begin{equation}
\mu(\d\xb) = \rho(\xb) \d\xb = Z^{-1} e^{-\beta U(\xb)} \d \xb,
\end{equation}
where $U:\RR^n\to \RR$ is the energy function, $\beta = 1/(k_{\rm B} T)$ is the inverse temperature in units of energy, and $Z$ is the canonical partition function.
We define a coarse-graining map
\begin{equation}
  g(\xb): \RR^n \to \RR^k \qquad g: \xb \mapsto \zb
\end{equation}
and emphasize again that, in general, we make no assumptions on the linearity of this map, though our numerical results use a linear coarse-graining map.
The marginal probability density for the coarse-grained variable is given by the conditional expectation, 
\begin{equation}
  \label{eq:marginal}
  e^{-\beta F(\zb)} = \EE_{\xb\sim \mu} \left[ 1 | g(\xb) = \zb \right] = Z^{-1} \int e^{-\beta U(\xb)} \delta(g(\xb)-\zb) \d \xb.
\end{equation}
The integral~\eqref{eq:marginal} cannot be computed directly.
The moments of the distribution can be accessed by numerically imposing the constraint $g(\xb)=\zb$ and evaluating conditional expectations.

Because we cannot easily compute the negative log-likelihood of the marginal probability density $F:\RR^k\to\RR$, also known as potential of mean force, we instead work with its gradient. 
Note that 
\begin{equation}
  \nabla_{\zb} F(\zb) = -\beta^{-1} \frac{\nabla_{\zb} \int e^{-\beta U(\xb)}  \delta(g(\xb)-\zb) \d \xb}{\int e^{-\beta U(\xb)} \delta(g(\xb)-\zb) \d \xb} 
\end{equation}
To see the connection between this conditional expectation and the mean force, let us first lift the map $\tilde{g}:\RR^n\to \RR^n$ and make the canonical change of coordinates $\tilde{g}(\xb)=\boldsymbol{g}$. 
The first $k$ dimensions of $\tilde{g}$ coincide with the coarse-graining map $g$. 
The gradient operates on the Dirac $\delta$-function, i.e., 
\begin{equation}
  \nabla_{\zb} F(\zb) = -\beta^{-1} \frac{ \int e^{-\beta (\tilde{U}(\boldsymbol{g}) - k_{\rm B} T \log |J({\gb})|)} \nabla_{\zb} \delta(\boldsymbol{g}_{[1,\dots,k]}-\zb)  \d \boldsymbol{g}}{\int e^{-\beta U(\xb)} \delta(g(\xb)-\zb) \d \xb} 
\end{equation}
and using the fact that
\begin{equation}
\nabla_{\zb} \delta (\boldsymbol{g}_{[1,\dots,k]}-\zb) = \nabla_{\boldsymbol{g}_{[1,\dots,k]}} \delta (\boldsymbol{g}_{[1,\dots,k]} - \zb), 
\end{equation}
we obtain 
\begin{equation}
  \nabla_{\zb} F(\zb) = \frac{\int \nabla_{\boldsymbol{g}_{[1,\dots,k]}} (\tilde{U}(\boldsymbol{g})- k_{\rm B} T \log |J(\gb)|) e^{-\beta \tilde{U}(\boldsymbol{g}) - k_{\rm B} T \log |J(\gb)|)}  \delta(\boldsymbol{g}_{[1,\dots,k]}-\zb)  \d \boldsymbol{g}}{\int e^{-\beta U(\xb)} \delta(g(\xb)-\zb) \d \xb}. 
\end{equation}
The resulting conditional expectation can be interpreted as the mean force on the constrained degrees of freedom $g(\xb)$.
Our objective is to build a parametric representation of $\nabla_{\zb} F(\zb)$ and to compare the efficiency of distinct estimators for this quantity. 

\subsection{Representation}

Throughout, we assume that the effective potential of mean force is a conservative vector field. 
This assumption holds whenever the fine-grained system is sampled at thermal equilibrium. 
To ensure that the force vector field that we parameterize is conservative, we represent the forces for regression as $\nabla_{\zb} U_{\theta}(\zb)$. With this representation, the time evolution of the probability distribution is governed by the Fokker-Planck equation,
\begin{equation}
\partial_t \tilde\rho_t(\zb) = \nabla \cdot \left( \nabla U_{\theta}(\zb) \tilde\rho_t(\zb) \right) + \beta^{-1} \Delta \tilde\rho_t(\zb).
\end{equation} 
This, in turn, ensures that the stationary distribution satisfies,
\begin{equation}
\rho_{\theta}(\zb) \propto e^{-\beta U_{\theta}(\zb)} \approx e^{-\beta F(\zb)} + \textrm{const.}
\end{equation}
It should be noted that choosing a representation that requires differentiation to evaluate can lead to scaling challenges.
These challenges are made especially acute when training using score-matching.

\section{Force Matching}

Force-matching is a widely used strategy for optimizing representations of the coarse-grained potential of mean force.
In its simplest form, force-matching minimizes
\begin{equation}
  \label{eq:fmloss}
  \mathcal{L}_{\rm FM}(\theta) = \EE_{\xb\sim \mu} \| - \nabla_{\zb} U(\xb) + \nabla_{\zb} U_{\theta} (g(\xb)) \|^2,
\end{equation}
where the components of the gradient of the atomistic potential are given by, 
\begin{equation}
  [\nabla_{\zb} U(\xb)]_j = \sum_{i=1}^n \diff{U}{x_i}{} \diff{x_i}{z_j}{} \implies \nabla_{\zb} U(\xb) = \nabla_{\xb} U \cdot \nabla_{\zb} \xb.
\end{equation}
The derivatives that we need to compute in this expression must be constructed implicitly because $\xb$ is not represented directly as a function of $\zb$.
To express this derivative, we note that, via the inverse function theorem,
\begin{equation}
\nabla_{\zb} g^{-1}(\zb) = |\nabla_{\xb} g(\xb)|^{-1} \qquad \textrm{for } g(\xb) = \zb.
\end{equation}
As a result, we see that the projected forces can be computed as
\begin{equation}
  \label{eq:projfs}
\nabla_{\xb} U \cdot (J[g])^{-1}(\xb)
\end{equation}
where $J^{-1}$ is the inverse of the Jacobian of $g$ evaluated at $\xb$.
Because $g$ is emphatically not an invertible map---it is a dimensionality reduction---the matrix we must invert is an element of $\RR^{k\times n}$ and hence we require a pseudoinverse in $\RR^{n\times k}$.

In practice, we estimate the loss function~\eqref{eq:fmloss} on a finite number of Boltzmann distributed samples from the high-dimensional ensemble. The estimator for the force-matching loss is,
\begin{equation}
\hat{\mathcal{L}}_{\rm FM}^{(N)}(\theta) = \frac{1}{N} \sum_{i=1}^N \left\| -\nabla_{\xb} U \cdot (J[g])^{-1}(\xb_i) + \nabla_{\zb} U_{\theta}(g(\xb_i)) \right\|^2, \quad \{\xb_i\} \sim \mu.
\end{equation}
This holds for an arbitrary nonlinear coarse-graining map $g.$
When the coarse-graining map is a linear function, 
\begin{equation}
  g(\xb) = A \xb
\end{equation}
for $A\in \RR^{k\times n}$, we need $A^{+}$, the Moore-Penrose pseudoinverse of $A$
\begin{equation}
 A^{+} = (A^T A)^{-1} A^T,
\end{equation}
to compute the projected forces \eqref{eq:projfs}.
As a result, the estimator of the force-matching loss becomes,
\begin{equation}
 \hat{\mathcal{L}}_{\rm FM}^{(N)}(\theta) = \frac{1}{N} \sum_{i=1}^N \left\| - A^{+}\nabla_{\xb} U + \nabla_{\zb} U_{\theta}(g(\xb_i)) \right\|^2, \quad \{\xb_i\} \sim \mu.
\end{equation}

\subsection{Mitigating Noise in Force Matching}

The force matching loss function uses noisy labels: the instantaneous estimates of the mean force~\eqref{eq:projfs} contribute the overall loss and complicate the learning objective. 
To characterize this, following~\cite{wang_multi-body_2021}, we examine a conventional bias-variance decomposition of the objective. 
Using the shorthands $f(\xb)$ for the projected force, $f_{\theta}(\zb)$ for the model force, and $\bar f(\zb)$ for the ground truth mean force, 
\begin{equation}
  \label{eq:fmloss_decomp}
\begin{aligned}
  \mathcal{L}_{\rm FM}(\theta) &= \EE_{\xb\sim \mu} \| f(\xb) - f_{\theta}(g(\xb)) \|^2, \\ 
  &= \EE_{\xb\sim \mu} \| f(\xb) - \bar f(g(\xb)) + \bar f(g(\xb)) - f_{\theta}(g(\xb)) \|^2, \\ 
  &= \EE_{\xb\sim \mu} \| f(\xb) - \bar f(g(\xb)) \|^2 + \|\bar f(g(\xb)) - f_{\theta}(g(\xb)) \|^2 \\+ &2\left( f(\xb) - \bar f(g(\xb)) \right)^T \left( \bar f(g(\xb)) - f_{\theta}(g(\xb)) \right).\\ 
\end{aligned}
\end{equation}
The cross term in this expression vanishes in expectation because 
\begin{equation}
  \begin{aligned}
    &\EE_{\xb\sim \mu} \left( f(\xb) - \bar f(g(\xb)) \right)^T \left( \bar f(g(\xb)) - f_{\theta}(g(\xb)) \right)\\
    & = \EE_{\xb\sim \mu} \EE_{\zb = g(\xb)} \left( f(\xb) - \bar f(\zb) \right)^T \left( \bar f(\zb) - f_{\theta}(\zb) \right)
  \end{aligned}
\end{equation}
and the first term vanishes under the conditional expectation.

We can identify the two non-vanishing terms in the loss as quantifying (1) the heteroscedastic error associated with single-sample estimates of the mean force and (2) the error in the representation of this mean force.
Explicitly,
\begin{equation}
  \label{eq:fmloss_noise_error}
  \mathcal{L}_{\rm FM}(\theta) = \EE_{\xb\sim \mu} \underbrace{\| f(\xb) - \bar f(g(\xb)) \|^2}_{\textrm{noise}} + \underbrace{\|\bar f(g(\xb)) - f_{\theta}(g(\xb)) \|^2}_{\textrm{error}}.
\end{equation}
It should be noted that the noise term arises in part from the choice of coarse-graining map $g$, vanishing in the limit that $g = \textrm{Id}$.
Ref.~\cite{kramer_statistically_2023} explores reparameterizations of $g$ to mitigate noise, but in this work we choose to fix the map to most clearly compare distinct training objectives and potential representations.

The noise term varies with the coarse-grained coordinate $z$ and mitigating it requires sampling a sufficient number of fine-grained configurations that the effective sample size at $z \gg 1$.
To see this, we define
\begin{equation}
    \delta(\xb) = f(\xb) - \bar{f}(g(\xb))
\end{equation}
so that the noise at finite sample numbers is
\begin{equation}
\textrm{var}^{(N)}(\zb) = \frac1N \sum_{i=1}^N \delta(\xb_i)^2, \quad \xb_i \sim \mu(\cdot|\zb).
\label{eq:condvar}
\end{equation}
This expression reveals that the noise controlled by the number of samples $\xb_i$ such that $g(\xb_i)=\zb.$
Because the fine-grained data is often generated using MD, the time autocorrelation in position yields configurations that map to similar values of $\zb$ provided that $g$ is a sufficiently smooth function of $\xb.$

Using correlated data with low signal-to-noise ratio ameliorates the variance in force matching, but can be counterproductive because it increases training cost.
A strikingly simple solution reduces both the noise and number of data points required to estimate the potential of mean force. 
Increasing $N$ in the sample variance \eqref{eq:condvar} controls the statistical fluctuations from Boltzmann sampling and suggests instead a ``mean force matching'' loss, 
\begin{equation}
  \label{eq:mfmloss}
  \mathcal{L}_{\rm MFM}(\theta) = \EE_{\xb\sim \mu} \left\| - \EE [ \nabla_{\zb} U(\xb) | g(\xb)=\zb] + \nabla_{\zb} U_{\theta} (\zb) \right\|^2,
\end{equation}
where the conditional expectation admits the straightforward estimator,
\begin{equation}
    \hat\EE^{(N)}_{\zb} [\nabla_{\zb} U(\xb)| g(\xb)=\zb] = \frac1N \sum_{i=1}^{N} \nabla_{\zb} U(\xb_i), \quad \xb_i \sim \mu(\cdot| g(\xb) = \zb).
\end{equation}
To collect samples conditioned on a coarse-grained variable $\zb,$ we perform constrained atomistic MD simulations. 

The impact of this change is significant, particularly for the noise term.
Indeed, with mean force matching, the bias-variance decomposition eliminates the noise term entirely and only finite $N$ estimator variance remains.
The significant practical advantage of this strategy is also apparent: mitigating noise in force matching requires autocorrelated samples, which increases the size of the data set.
With mean force matching, the coarse-grained coordinate can be sampled iid from any suitable distribution, improving coverage of the CG landscape. 

\subsection{Evaluating Score Matching}

Hyv\"arinen score matching~\cite{hyvarinen_estimation_2005} provides an alternative that does not require access to microscopic forces at all.
This procedure directly learns the distribution of the data by regression of the gradients of the free energy, which have a self-consistent form.
Assuming that our data is distributed according to the target Boltzmann distribution with density $\rho$, the score matching seeks to minimize the objective, 
\begin{equation}
  \EE_{\rho} \| \nabla U_{\theta}(\zb) + \nabla \log \tilde{\rho}(\zb) \|^2,
\end{equation}
where $\tilde{\rho}$ denotes the probability density of the system projected onto the coarse-grained variables.
This expectation admits a tractable estimator because
\begin{equation}
  \begin{aligned}
  \EE_{\hat{\rho}} \| \nabla U_{\theta}(\zb) + \nabla \log \tilde{\rho}(\zb) \|^2 &= 
\EE_{\hat{\rho}} \|\nabla U_{\theta}(\zb) \|^2 + 2 \nabla U_{\theta}(\zb)\cdot \nabla \log \tilde{\rho}(\zb) + \textrm{const.} \\
&= \EE_{\hat{\rho}} \| \nabla U_{\theta}(\zb) \|^2 - 2 \Delta U_{\theta}(\zb) + \textrm{const.} 
  \end{aligned}
\end{equation}
where $\Delta$ denotes the Laplacian operator.

While score matching eliminates the noisy regression required by force matching, it brings additional complexity to the learning problem.
First, score matching strictly requires that the data distribution is Boltzmann distributed; this severely limits the training data mixtures and complicates the effort to include rare, high signal configurations.
Secondly, ensuring that the CG potential is a conservative vector field inhibits scalability because evaluating the Laplacian $\Delta U_{\theta}$ is both memory intensive and computationally costly.
To reduce the cost, we use a form of Stein's lemma~\cite{chen_normal_2010} that approximates the Laplacian using a contraction with an iid Gaussian random variable and a stochastic finite difference, 
\begin{equation}
    \label{eq:trace}
    \Delta U_{\theta}(\zb) \approx \frac{\epsilon^{-1}}{N_{\textrm{sample}}} \sum_{i=1}^{N_{\textrm{sample}}} \etab_i^T \left[ \nabla U_{\theta}(\zb + \epsilon \boldsymbol{\eta}_i) - \nabla U_{\theta}(\zb) \right],
\end{equation}
where $\etab \sim \mathcal{N}(0, \mathrm{Id})$ is a Gaussian random variable and $\epsilon>0$ is a parameter taken to be as small as numerical stability permits.
The additional variance in the loss from Hutchinson's trace estimator is quantified in Fig.~\ref{fig:trace}.
The memory costs of Hyv\"arinen score matching can also be ameliorated using denoising score matching~\cite{vincent_connection_2011}, which could be used as an auxiliary loss, similar to Ref.~\cite{arts_two_2023}.
However, obtaining good training signal requires long noising trajectories~\cite{NEURIPS2022_a98846e9} which results in incurring the cost of training a separate denoising diffusion model.

\section{Constructing a Benchmark for Transferable CG}

\subsection{Dataset Design}
We constructed datasets to enable direct comparisons of performance and cost across different training objectives. To ensure broad structural diversity, we selected initial structures from the mdCATH dataset \cite{mirarchi_mdcath_2024}. This dataset provides MD trajectories at several temperatures for nearly all families within the CATH system, a database that classifies protein domains into families based on structural and evolutionary relationship, helping to ensure broad structural coverage. For precise comparisons between training protocols, we used the same set of 20,000 starting structures across 1000 CATH domains sampled from mdCATH trajectories performed at 320 K. 

For training with the MFM objective, we aimed to build a dataset that approximates the true mean force of each CG configuration. This was accomplished by fixing atoms corresponding to our CG beads (C$\alpha$, C, and N of the backbone) and performing MD simulations with these constraints until the standard error in the atomistic forces were less than 1 $k_{\textrm{B}}T$ per CG bead. Forces were then averaged over the simulation to obtain mean force estimates. This procedure typically required 2-4 ns of simulation per configuration.
While these estimates may be biased due to slow relaxations in side chain conformations, as shown in the results below, this effect is strongly mitigated by decreased variance. Alternative protocols for preparing initial conditions could mitigate this concern, but we sought here to make the datasets as comparable as possible. 

For training with the FM and SM objectives, unrestrained simulations were performed using the same starting structures as used for MFM. For each domain, we sampled 1000 frames from 300-400 ns of simulation, resulting in a dataset with one million total data points.

\begin{figure}
    \centering
    \includegraphics[width=0.42\linewidth]{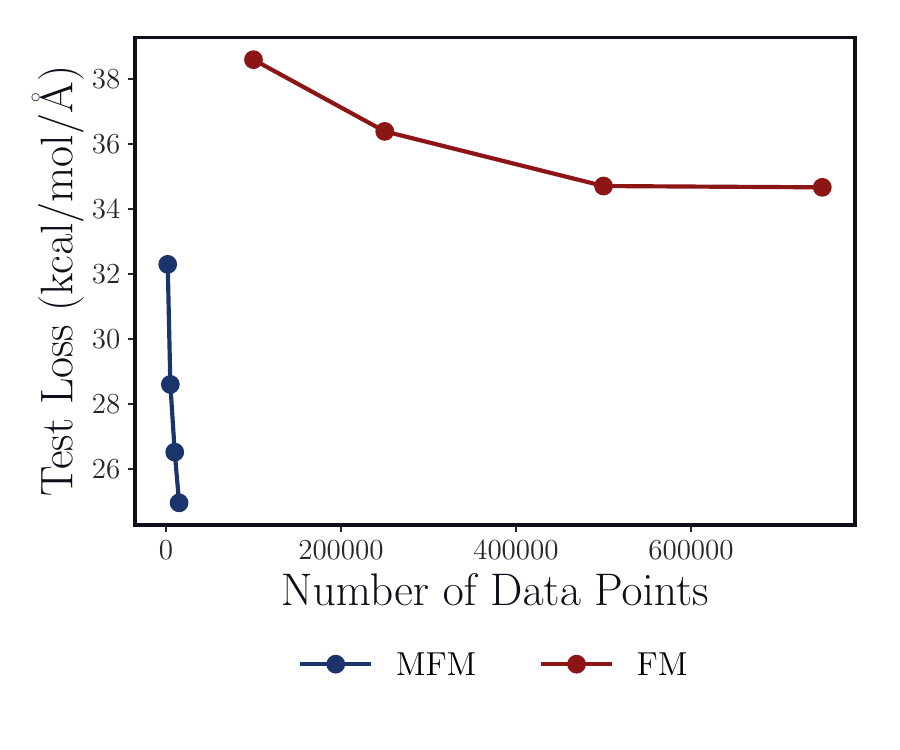}
    \includegraphics[width=0.56\linewidth]{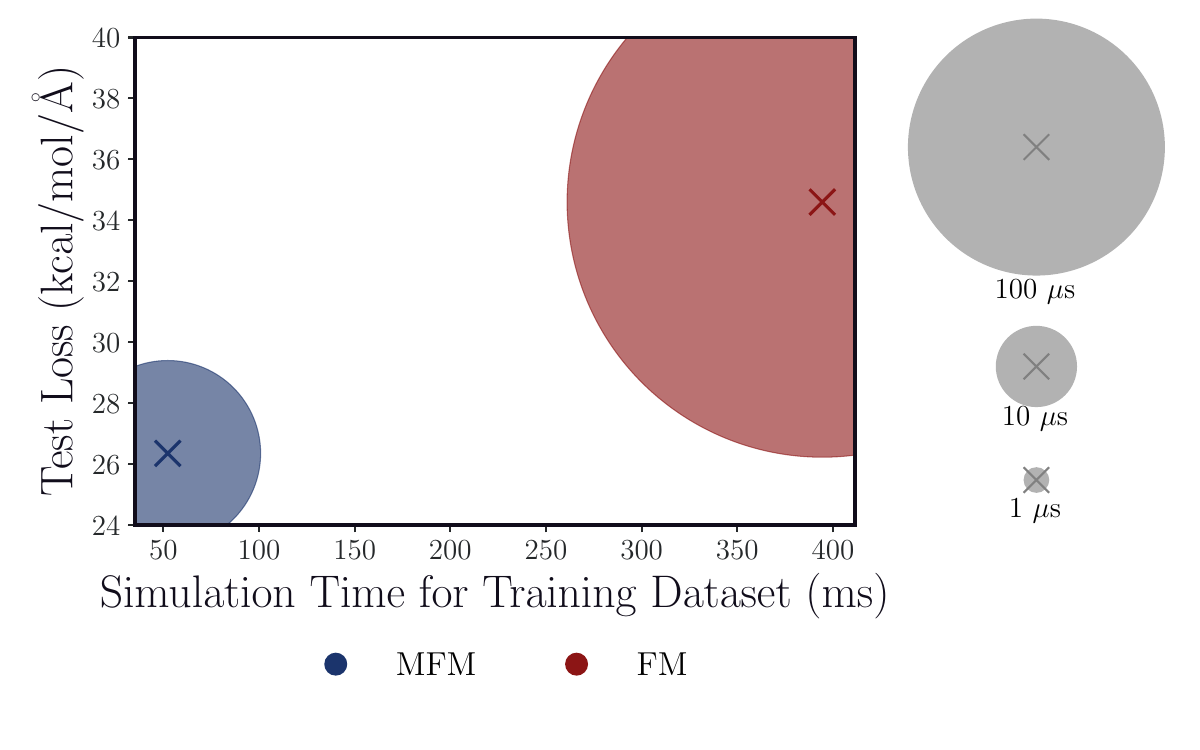}
    \caption{Data efficiency comparison of MFM and FM (a) Test loss scaling of MFM and FM on randomly sampled subsets of the full dataset (b) Test loss for MFM and FM as a function of total simulation time required for the full dataset. All models use the MACE architecture and are evaluated on mean force estimates from a held-out test of 50 CATH domains.}
    \label{fig:datascaling}
\end{figure}

As shown in Fig.~\ref{fig:datascaling}, increasing training dataset size improves performance for both the MFM and FM objectives. However, the MFM objective requires significantly less data than FM to achieve a given test loss, reducing data acquisition compute costs to a fraction of those for FM. Using the MACE architecture \cite{batatia_mace_2023}, models were trained on randomly sampled subsets of the full MFM and FM training datasets and evaluated on mean force estimates of a held-out test set of 50 CATH domains. Small increases in dataset size for training with the MFM objective substantially improved test performance, with the MFM model trained only on 2,000 data points achieving a lower test loss than the FM model trained on 750,000 data points, representing a 375-fold reduction in training data.

Leveraging MFM's data efficiency, we expanded the mean force dataset to 100,000 data points spanning the same 1000 CATH domains but with more diverse configurations per domain by sampling from higher temperature mdCATH simulations. We refer to this expanded dataset as MFM 100K in subsequent results.

\subsection{Benchmarking Architectures and Objectives}

\begin{figure}
    \centering
    \includegraphics[width=0.49\linewidth]{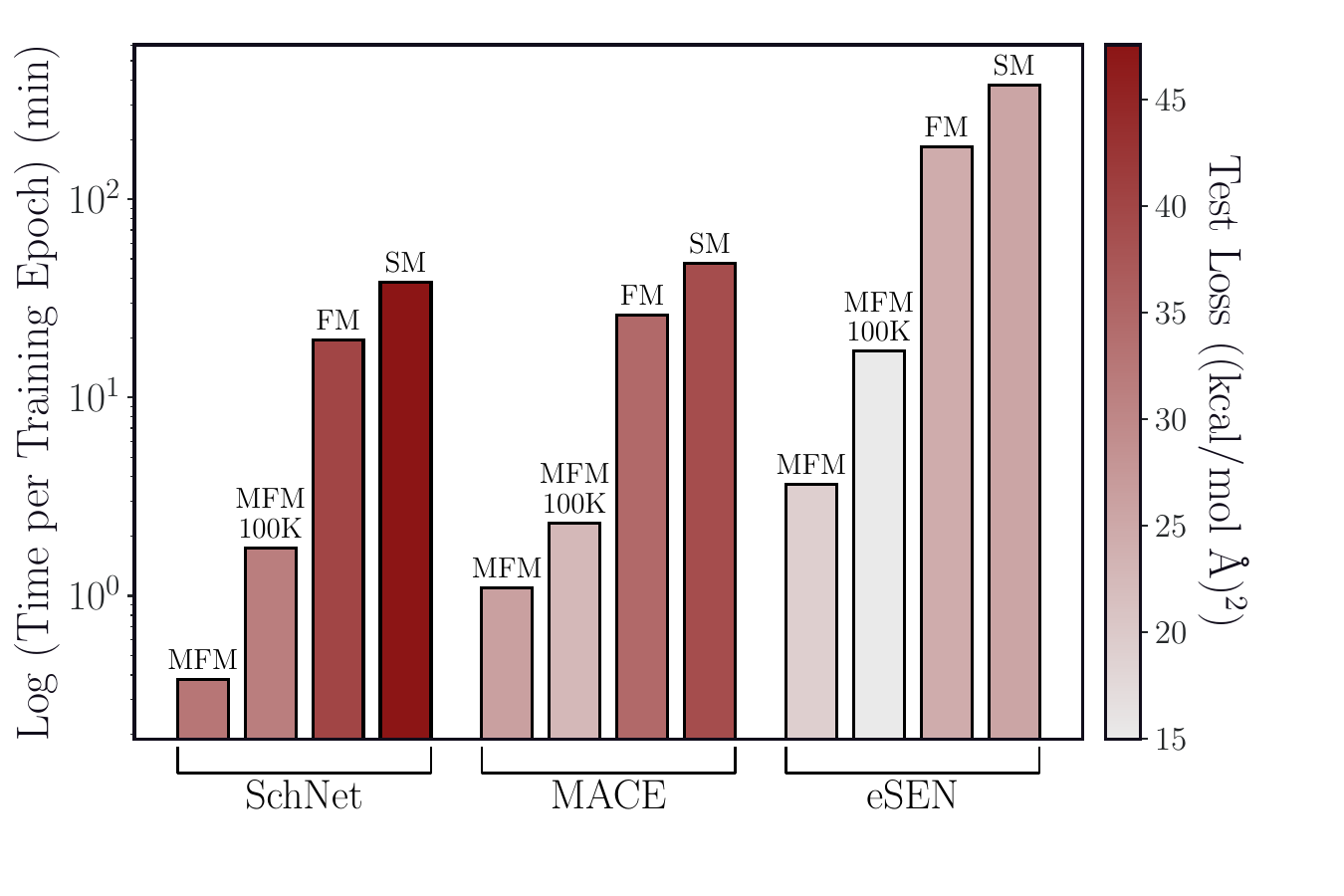}
    \includegraphics[width=0.49\linewidth]{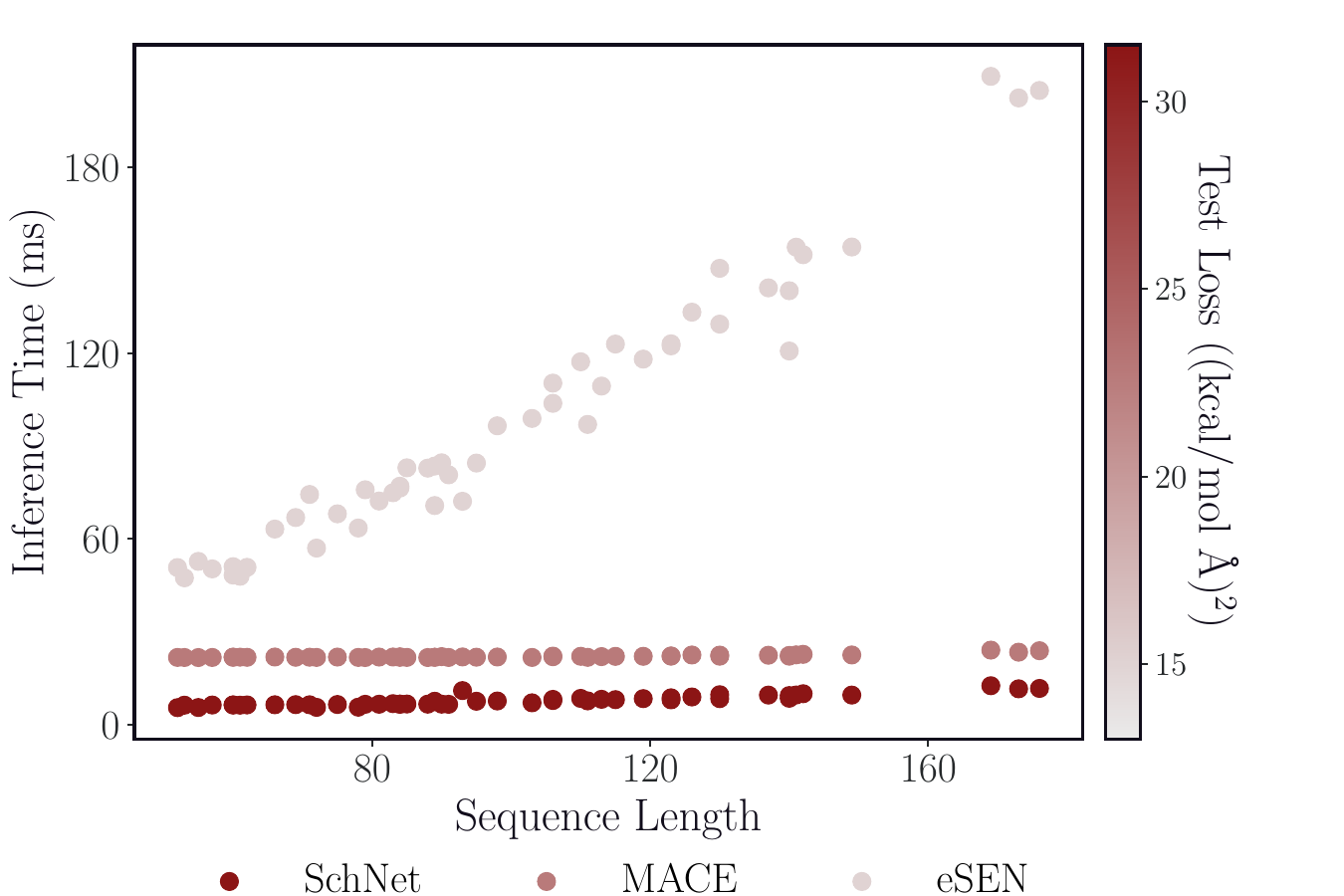}
    \caption {Computational costs of training and inference (a) Wall-clock time per training epoch of MFM, MFM 100K, FM, and SM objectives across SchNet, MACE, and eSEN architectures (b) Inference time as a function of protein sequence length for models trained via MFM 100K.}
    \label{fig:modelcomp}
\end{figure}

To assess how model expressiveness and scale impact performance, we trained several architectures, including SchNet \cite{schutt_schnet_2018}, MACE \cite{batatia_mace_2023}, and eSEN \cite{fu_learning_2025}, using each of the objectives discussed above. Although there are post-processing \cite{kramer_statistically_2023} and data augmentation \cite{durumeric_learning_2024} strategies that have been shown to modestly reduce variance of FM and SM, we did not apply these techniques and instead followed a standard protocol for the FM and SM objective. After training, we evaluated accuracy on a held-out test set of mean force estimates from configurations of 50 CATH domains that were not included in the training set. We generally find that mean-squared error (MSE) in mean force in a held-out test set correlates well with fidelity of thermodynamic observables, as discussed in Sec.~\ref{sec:results}.

In Fig.~\ref{fig:modelcomp}, we compare the performance, training costs, and inference-time cost of SchNet, MACE, and eSEN. Across all architectures, models trained with the MFM objective consistently achieved lower test loss than those trained with the FM or SM objectives. In particular, the eSEN model trained using MFM on the 100K dataset attained the lowest test loss of 14.89 kcal/mol \AA. However, the increased expressiveness of eSEN comes at a substantial computational cost, scaling poorly with system size. MACE offers a good balance between expressiveness and efficiency. Beyond improved test performance, MFM also proved much more efficient to train than FM and SM. Using the MACE architecture, a single training epoch with MFM on the 100K dataset was over 10 times faster than FM and 20 times faster than SM, portending further scalability with increased data.

\begin{table}[h]
\centering
\begin{tabular}{c|cccc}
    & MFM & MFM 100K & FM & SM \\ \hline
    SchNet & 32.74 & 31.50 & 40.11 & 47.59 \\[6pt]
    MACE & 26.35 & 22.74 & 34.60 & 38.94 \\[6pt]
    eSEN & 19.16 & 14.97 & 24.44 & 25.64 \\[6pt]
    \end{tabular}
    \caption{Test MSE across 50 CATH domains ((kcal/mol \AA)$^2$).}
\end{table} 

\section{Evaluating Zero-Shot Performance on Thermodynamic Observables}\label{sec:results}

While test loss indicates a clear trend in model performance, we computed the free energy surfaces (FES) of several well-characterized proteins that were not in our training set to assess out-of-domain generalization, which consists primarily of folded proteins. Specifically, we examined whether our models could accurately reproduce FES of fast-folding proteins, Trp-cage and BBA~\cite{lindorff-larsen_how_2011}.

\begin{figure}
    \centering
    \includegraphics[width=0.32\linewidth]{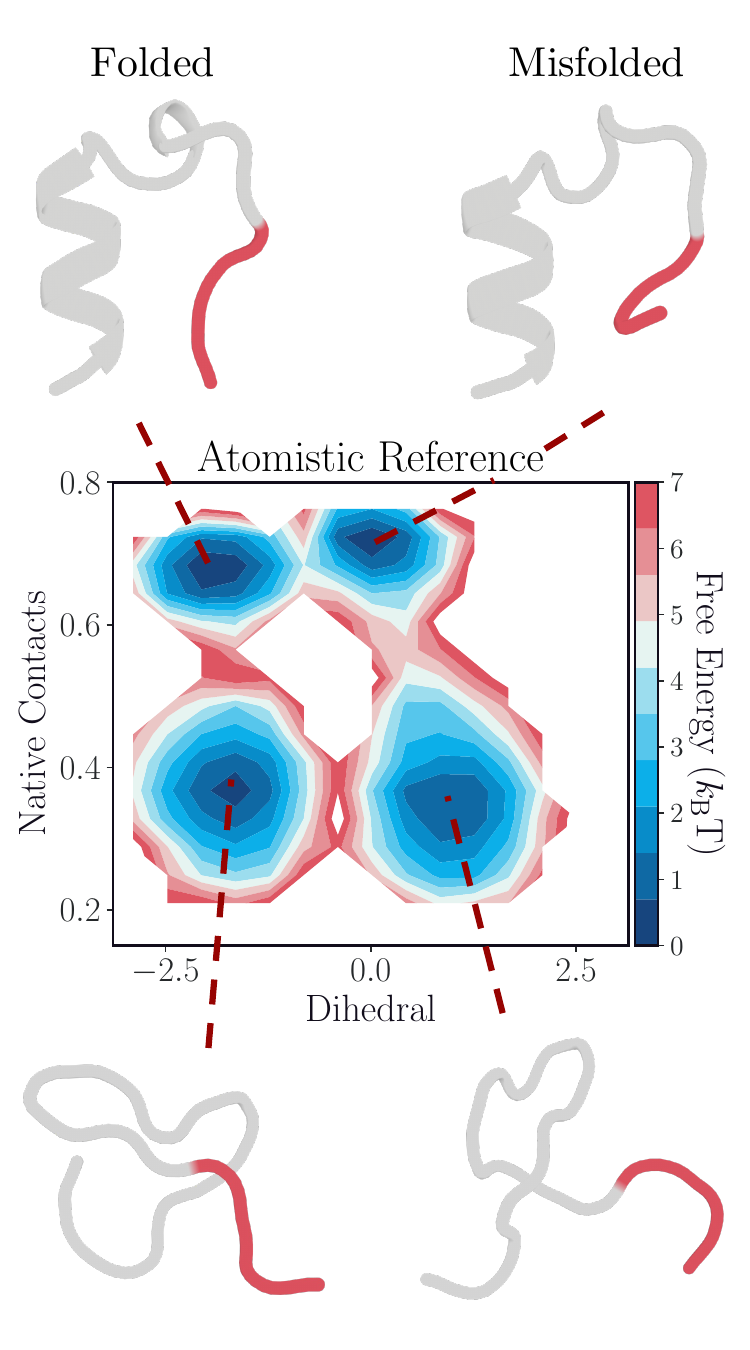}
    \includegraphics[width=0.64\linewidth]{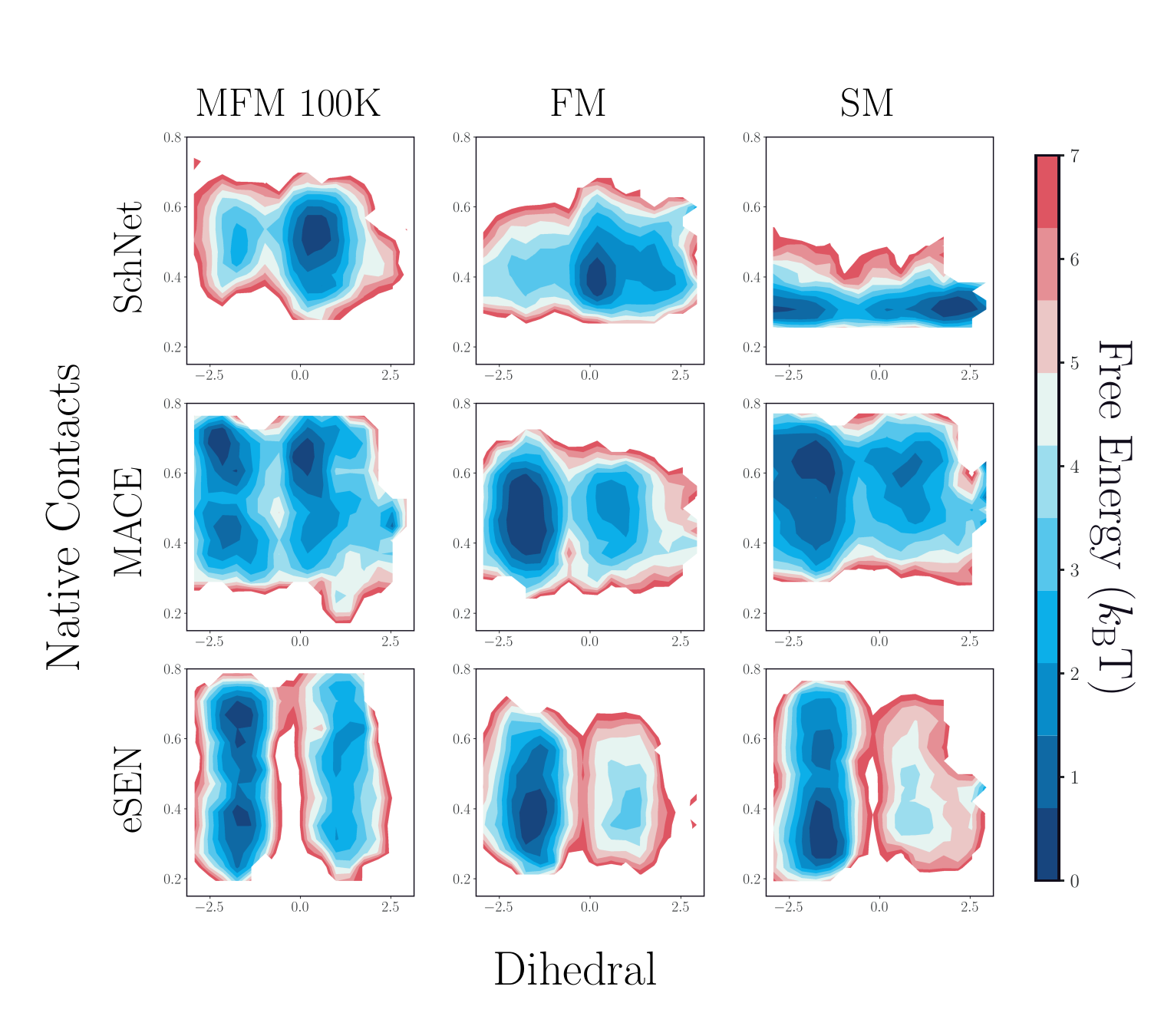}
    \caption{Free energy surface (FES) of Trp-cage (a) Atomistic reference FES derived from over 30 $\mu$s of explicit solvent MD at 300 K with representative structures from each metastable state. Dihedral of the last four alpha carbons (shown in red) was used for the first CV and fraction of native contacts as the second. (b) FES produced via umbrella sampling by CG models across different training objectives (MFM, FM, and SM) and model architectures (SchNet, MACE, eSEN)}
    \label{fig:trpcage}
\end{figure}

We first evaluated our CG models on Trp-cage, a 20 residue protein featuring a folded state with an alpha helix and a hydrophobic ``cage'' around a tryptophan residue \cite{gelman_fast_2014}. The atomistic reference was generated by performing over 60 $\mu$s of simulation at 300 K in explicit solvent. Using the dihedral angle between the last four alpha carbons and fraction of native contacts as collective variables (CVs), the atomistic reference in Fig.~\ref{fig:trpcage} reveals four metastable states: a folded, misfolded, and two distinct unfolded states. 

\begin{figure}
    \centering
    \includegraphics[width=\linewidth]{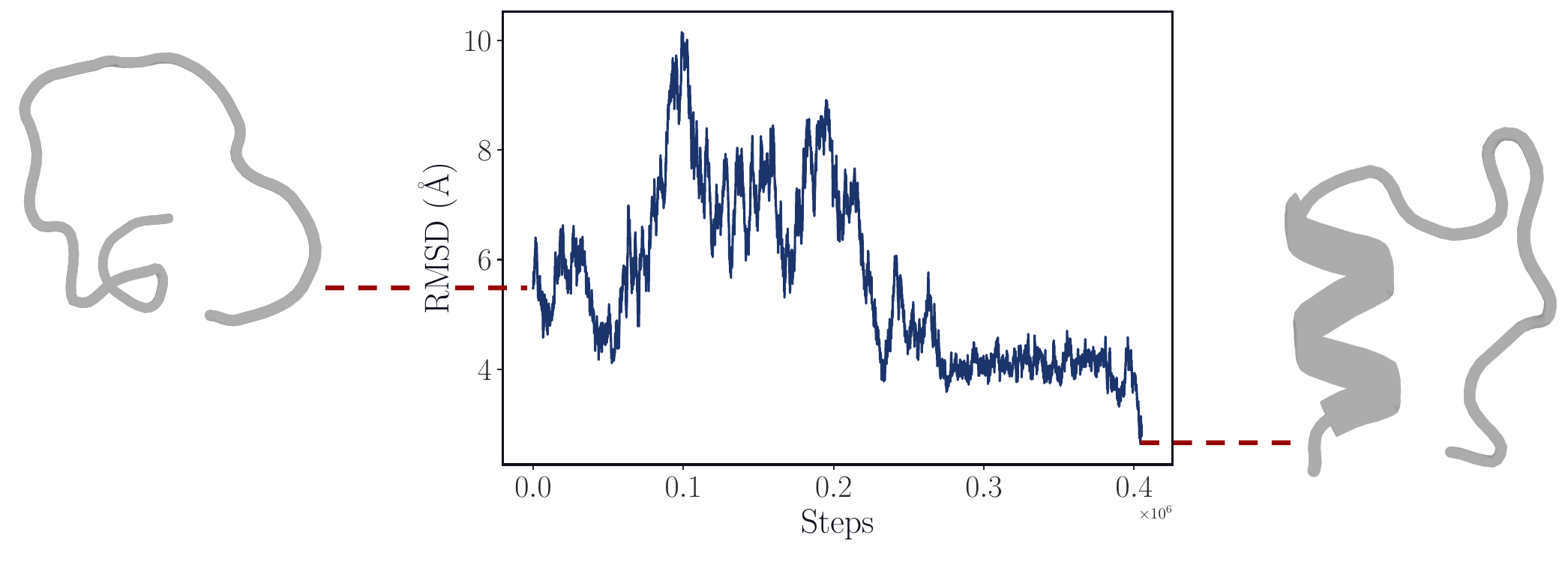}
    \caption{RMSD to native conformation of folding trajectory of Trp-cage starting from unfolded configuration}
    \label{fig:trp_folding}
\end{figure}

FES calculations for our CG models were performed using umbrella sampling. Regardless of the training objective SchNet models performed poorly, failing to stabilize the folded state or distinguish between the metastable states. In contrast, the MACE and eSEN models trained via MFM on the 100K dataset reproduced essential features of the atomistic reference, capturing, in particular, the relevant metastable states. Models trained via FM and SM identified the two dihedral states, but failed to reliably differentiate between folded and unfolded states based on native contacts. Further, unbiased Langevin simulations using the MACE MFM 100K model successfully folds Trp-cage from an unfolded configuration as shown in Fig.~\ref{fig:trp_folding}.

\begin{figure}
    \centering
    \includegraphics[width=0.32\linewidth]{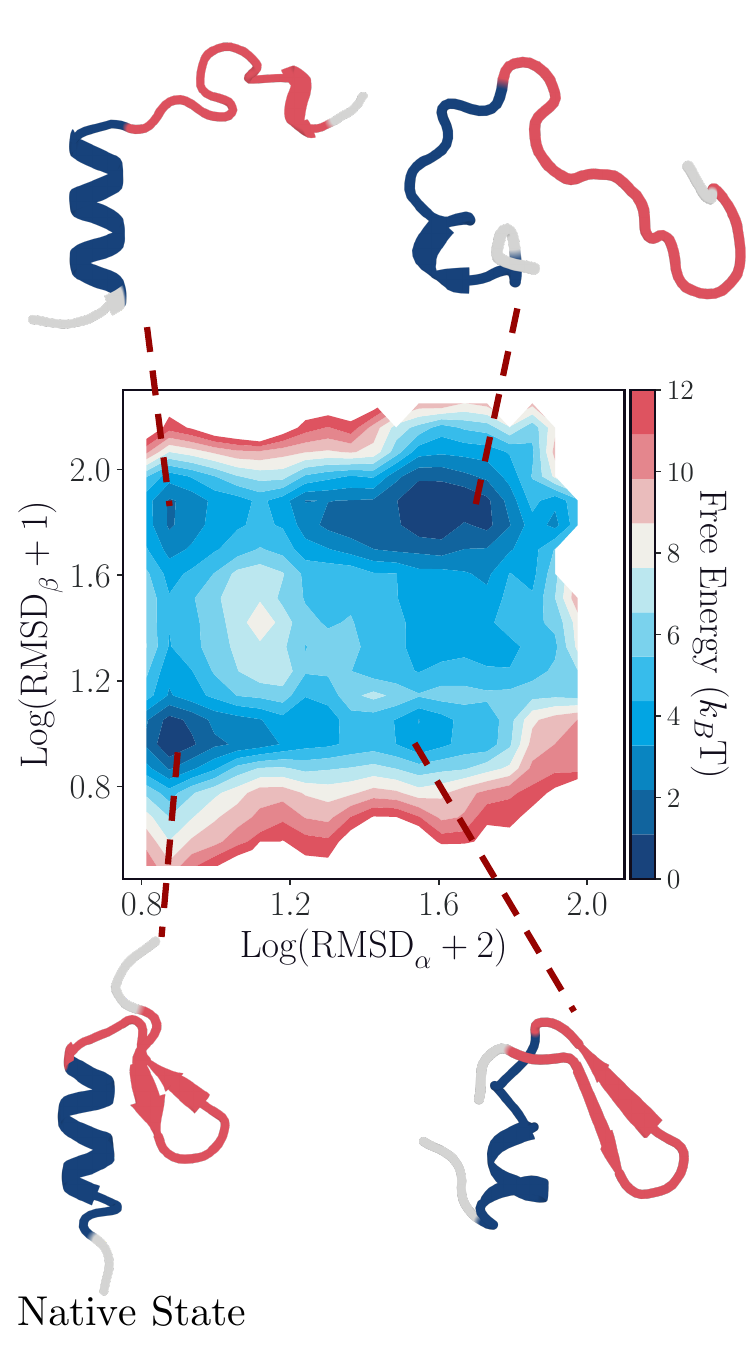}
    \includegraphics[width=0.64\linewidth]{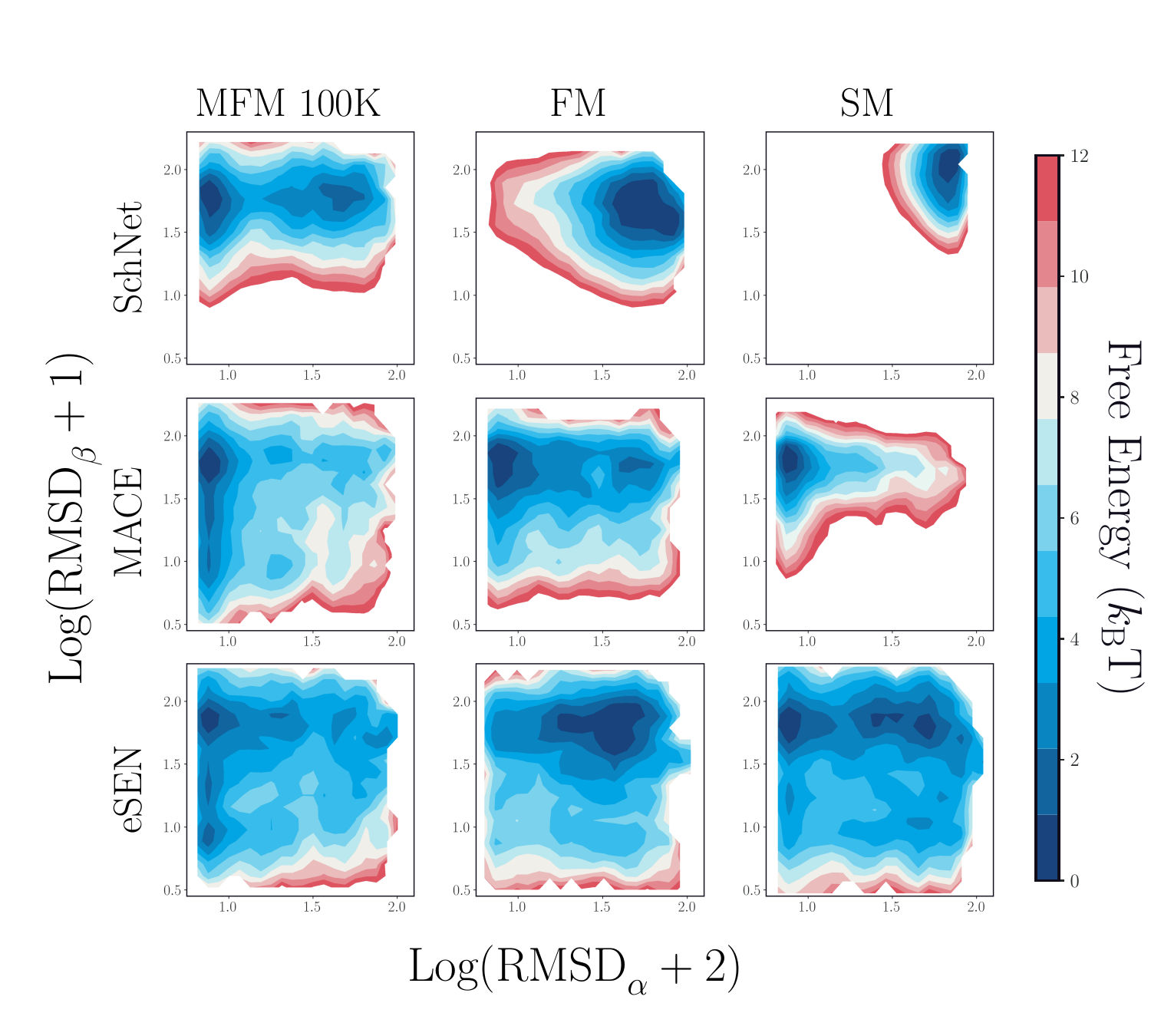}
    \caption{Free energy surface of BBA (a) Atomistic reference FES produced through umbrella sampling in explicit solvent at 300 K with representative structures from each metastable state. RMSD$_{\alpha}$ refers to RMSD of residues 16-26 (shown in red) to an ideal $\alpha$-helix. RMSD$_{\beta}$ refers to RMSD of residues 4-15 (shown in blue) to native state  (b) FES produced via umbrella sampling by CG models across different training objectives (MFM, FM, and SM) and model architectures (SchNet, MACE, eSEN)}
    \label{fig:bba}
\end{figure}

We observed similar trends for BBA, a 28 amino-acid protein based on the zinc finger $\beta\beta\alpha$ motif, containing both an $\alpha$-helix and $\beta$-hairpin \cite{gelman_fast_2014}. For this system, umbrella sampling employed two CVs: the first based on the RMSD of the $\alpha$-helical region to an ideal reference structure and the second based on the RMSD of the $\beta$-hairpin region to its native conformation. The atomistic reference in Fig.~\ref{fig:bba} exhibits four states: fully folded, folded helix with unfolded hairpin, unfolded helix with folded hairpin, and fully unfolded.

The MFM 100K models again demonstrated superior performance across architectures. Although no SchNet model stabilized the $\beta$-hairpin, the SchNet MFM 100K model was able to stabilize the helix, producing a FES with two basins corresponding to a folded helix/unfolded hairpin state and a fully unfolded state. Among the MACE models, the MFM 100K model closely resembled the atomistic reference, stabilizing both structural elements with a basin corresponding to the native state. The MACE FM and SM models stabilized the helix, but not the hairpin. Similarly, the eSEN MFM 100K model successfully captured the hairpin and helix, with its FES showing a fully folded basin. The eSEN FM model failed to stabilize the helix or hairpin, but the SM model's FES showed basins corresponding to a folded helix and folded hairpin. With maximum sequence similarities of 50\% for Trp-cage and 42.9\% for BBA to the training data, MFM demonstrates robust zero-shot transferability by producing the most accurate FES for proteins outside the training domain.

\section{Comparisons with Folded-state Atomistic MD}
\begin{figure}
    \centering
    \includegraphics[width=\linewidth]{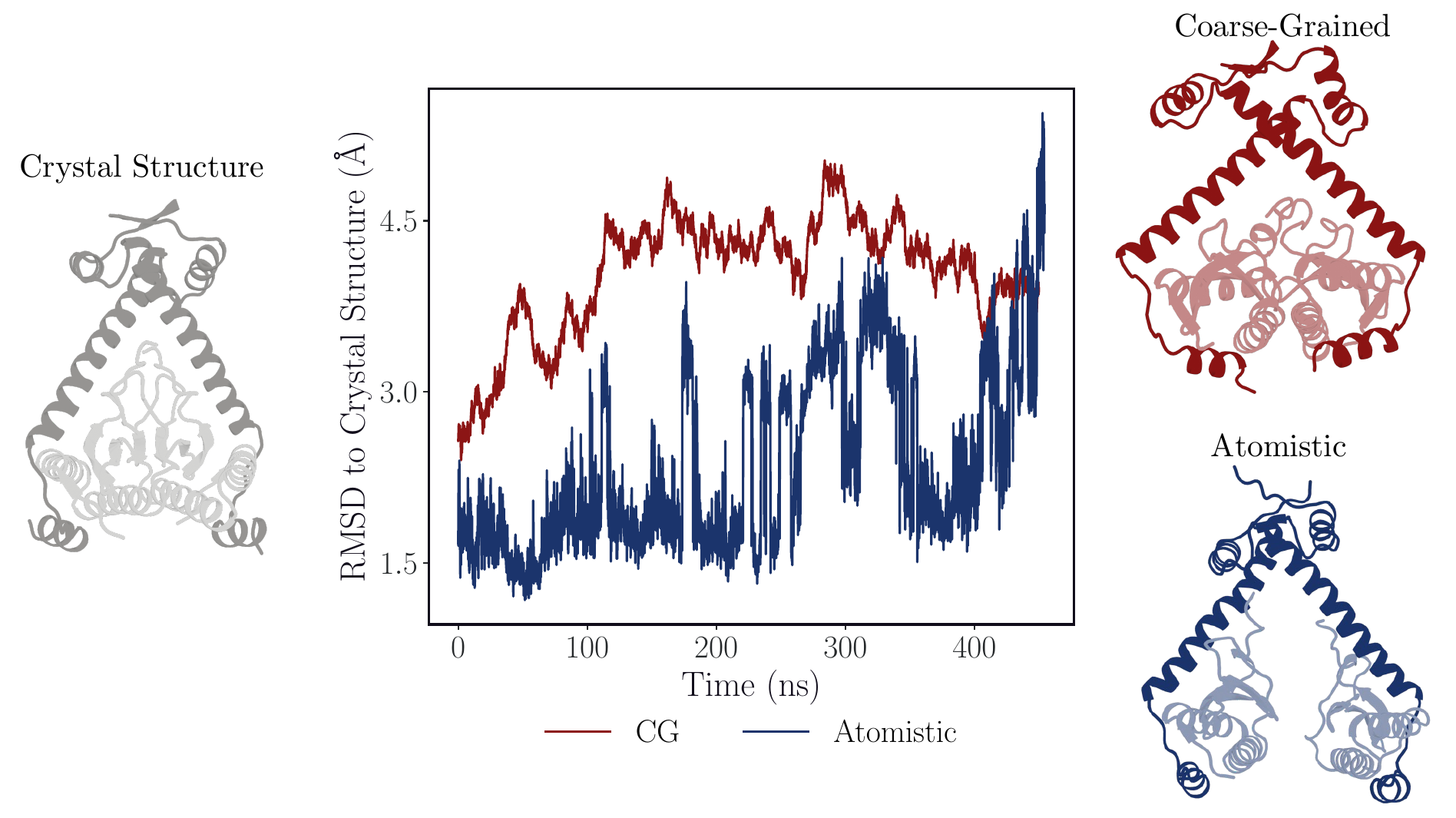}
    \caption{RMSD to crystal structure of atomistic and CG simulations of ParE-ParD complex. Final frame of the CG (red) and atomistic (blue) simulation shown on right.}
    \label{fig:rmsd_pare_pard}
\end{figure}

\begin{figure}
    \centering
    \includegraphics[width=\linewidth]{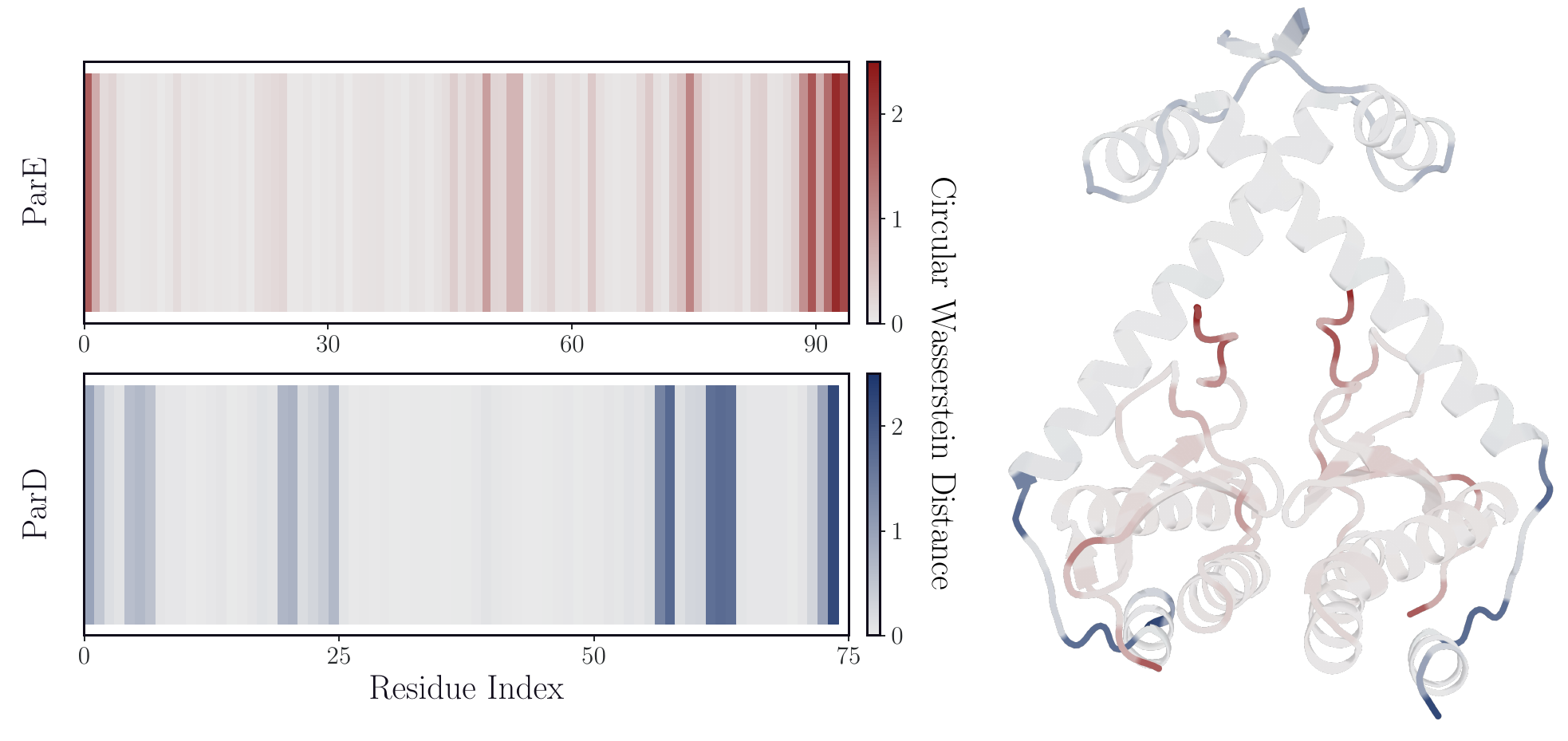}
    \caption{(a) Circular Wasserstein-1 Distance between backbone dihedral distributions of atomistic and CG simulations (b) Crystal structure colored by per-residue error}
    \label{fig:wasserstein}
\end{figure}

While our model was trained exclusively on single-chain, monomeric folded proteins, we assessed our CG model's generalization to protein complexes. We evaluated performance on the ParD-ParE toxin-antitoxin complex, an $\alpha_2\beta_2$ heterotetramer \cite{dalton_conserved_2010} with less than 40\% maximum sequence identity to the training set. We performed CG simulation using the MACE MFM 100K model starting from the crystal structure and compared the results to atomistic MD in explicit solvent. As shown in Fig. \ref{fig:rmsd_pare_pard}, the CG model exhibited similar RMSD to the crystal structure as the atomistic reference after 450 ns of simulation. To further assess our CG model's agreement with the atomistic reference, we considered the dihedral distributions of the backbone. Fig. \ref{fig:wasserstein} shows the dihedral distribution error between the CG model and atomistic reference as circular Wasserstein-1 distance per residue. The CG model shows low error in the highly structured regions of the complex, indicating that secondary structure remains stable under our CG model. Higher error corresponds to flexible loop regions, where the CG model explores a broader range of backbone conformations. These results demonstrate our CG model's strong generalization outside its training domain.




\section{Conclusions}

Through both systematic theoretical analysis and exhaustive numerical experiments, we show here that mean force matching yields results that dramatically outperform alternative objective functions for bottom-up coarse-graining. 
Because mean force matching reduces both noise in the objective function and also the data requirements, this objective also provides a significant advantage for scaling training of a CG potential to hundreds of proteins with highly expressive neural networks.
There are, of course, restrictions and limitations that arise from our approach: much of the computational effort arises in the data acquisition phase, which requires fixing the coarse-graining map. 
As a consequence, changing the map requires recomputing all training data, so we view this approach as being more suited to intuitive and physically motivated dimensionality reductions, as opposed to learned CG maps. 
Nevertheless, we find that the generalization afforded by mean force matching makes a compelling case for our approach.

Additionally, while scaling MLIPs produces quantitatively better results, inference comes with higher computational and memory costs in neural networks with large parameter counts, motivating further exploration of architectures suitable for coarse-graining.
We see this most dramatically with eSEN, which yields the best performance on test loss, but scales poorly with input protein size.
While MACE strikes a good balance between cost and accuracy, the gap to existing coarse-grained models in cost remains large. 
Designing architectures that directly incorporate inductive biases suitable for coarse-grained representation of proteins, such as local many-body effects and renormalized long-ranged interactions, may help close this gap. 
The data and experiments we describe here form an excellent benchmark for future architecture developments because rapid iteration on architecture design would be significantly more difficult using data intensive objectives like force matching or score matching. 

We focused on assessing ``zero-shot'' generalization on the free energy surfaces of a number of proteins, which show strong qualitative agreement, but system specific adaptation of our transferable potential would improve this even further.
Indeed, our results are sufficiently transferable that it will likely prove productive to treat this model as a foundation model that can be fine-tuned using forces from a system of particular interest.
Furthermore, specialization offers a plausible route to reducing computational costs even further by distilling to a smaller, more efficient model~\cite{amin2025distill}. 

While general purpose pre-trained MLIPs that leverage high quality electronic structure data are finding use as an alternative to \emph{ab initio} MD, there are not yet widely used foundation models for biomolecular thermodynamics and dynamics.
Though purely data-driven generative models, such as BioEmu~\cite{lewis_scalable_2024}, are being developed with this goal in mind, our approach prioritizes a strong physical prior and relies more on computation to generate samples.
However, we foresee incorporating experimental measurements of protein stability and kinetics to further improve model performance as the scale and accuracy of our models grows. 

\paragraph{Acknowledgements} This material is based upon work supported by the National Science Foundation under grant number CHE-2441297. This work used resources from the National Energy Research Scientific Computing Center under award number BES-ERCAP0033250. The authors thank Eric Vanden-Eijnden for useful discussions on Hutchinson's trace estimator.  

\bibliographystyle{unsrt}
\bibliography{references}

\clearpage
\newpage
\section{Supplementary Information}
\subsection{Dataset Construction}
\subsubsection{Initial Structure Preparation}
Initial structures were sampled from the mdCATH dataset~\cite{mirarchi_mdcath_2024} for 1000 CATH domains. The mdCATH dataset comprises 5 replica simulations, each extending 400 ns across five temperatures, 320 K, 348 K, 379 K, 413 K, and 450 K, yielding 25 simulations per domain. To select our initial configurations, we randomly sampled four frames from each of the 5 simulations conducted at 320 K, for a total of 20 starting configurations per CATH domain. Each structure was solvated using the TIP3P water model, with a 9 \AA\ cubic buffer at 150 mM salt concentration. All atomistic simulations were performed using the AMBER ff99SBdisp~\cite{robustelli_developing_2018} force field and OpenMM (version 8.1.1), with the OVRVO integrator~\cite{leimkuhler_rational_2012} at 300 K, a time step of 2 fs, and friction coefficient of 1.0 $\textrm{ps}^{-1}$.

\subsubsection{Atomistic Simulations for FM and SM datasets}
Each of the 20 initial configurations per CATH domain underwent a 5 ns equilibration period followed by a 20 ns production run. Trajectory frames were saved every 10 ps, yielding 2,000 frames per production simulation. Each CATH domain completed at least 15 production runs, providing 300–400 ns of aggregate simulation per domain. For training the FM and SM models, 1000 frames were randomly sampled from the combined trajectories of each CATH domain.

\subsubsection{Mean Force Calculations}
Constrained simulations were performed to estimate the mean force acting on the coarse-grained (CG) beads of each atomistic configuration. Atoms corresponding to CG bead sites (backbone $\textrm{C}\alpha$, C, and N atoms) were held fixed by setting their masses to zero. Trajectory frames were saved every 1 ps, and after an initial 1 ns period, the standard error of the mean force was computed at 1 ps intervals. Simulations were terminated once the standard error per bead fell below 1 $k_{\textrm{B}}T$, at which point the time-averaged mean force was recorded. The same 20 initial configurations per CATH domain used for the FM and SM datasets were used for the mean force dataset. Each domain completed at least 15 mean force estimates, yielding a total of 19,721 estimates for training the MFM model.

The expanded MFM 100K dataset included initial structures sampled from the higher temperature mdCATH trajectories. Spanning the same 1000 CATH domains, an additional 20 structures were sampled from the 348 K, 379 K, 413 K, and 450 K mdCATH simulations. Mean forces were estimated for these additional structures at 300 K following the same procedure as described above. Each CATH domain completed at least 90 mean force estimates (including those from structures sampled at 320 K). This dataset included 97,541 estimates.

The test set consisted of structures from 50 held-out CATH domains, which each had sequence similarity less than 40\% to the training dataset. Mean forces were estimated for 100 structures sampled from the 320 K, 348 K, 379 K, 413 K, and 450 K mdCATH simulations. Each CATH domain retained at least 90 mean force estimates, yielding a total of 4,885 mean force estimates.

\subsection{CG Model}
\subsubsection{Model Architecture}
Our CG models take the form of 
$$U_{\theta}^{\textrm{CG}} (R) = U_{\theta}^{\textrm{NN}}(R) + U_{\theta}^{\textrm{Prior}}(R)$$
where $U_{\theta}^{\textrm{NN}}$ was either SchNet, MACE, or eSEN and the prior resembled a classical MD forcefield. The prior took the form of,
\begin{align*}
    U_{\theta}^{\textrm{Prior}}(R) &= \sum_{\textrm{bonds}}k_{\textrm{bond}_i}(r_{i} - \mu_{\textrm{bond}_i}) + \sum_{\textrm{angles}} k_{\theta, \textrm{angles}_i}(\phi_i - \mu_{\theta, \textrm{angles}_i}) \\ 
    &+\sum_{\textrm{dihedrals}}\sum_{x=1}^{10} (A_{\theta, \textrm{dihedral}_i, x}\cos(x \psi_{i}) + B_{\theta, \textrm{dihedral}_i, x}\sin(x \psi_i)) \\ 
    &+ \sum_i^{N-4} \sum_{j=i+4}^N \epsilon_{\theta, i} \epsilon_{\theta, i}[(\frac{\sigma_{\theta, j} + \sigma_{\theta, j}}{r_{ij}})^{12} - (\frac{\sigma_{\theta, i} + \sigma_{\theta, j}}{r_{ij}})^{6}] + \frac{q_{\theta, i}q_{\theta, j}}{r_{ij}} \\ 
\end{align*}
All parameters with subscript $\theta$ are trainable in our implementation. 
The SchNet architecture was implemented based on the description in Ref. \cite{schutt_schnet_2018}. The MACE architecture was implemented by modifying code from \url{https://github.com/ACEsuit/mace}. The eSEN architecture was implemented by modifying code from \url{https://github.com/facebookresearch/fairchem}. 

\subsubsection{Prior Pre-training}
The prior was first pre-trained on the tripeptide dataset from Ref.~\cite{chennakesavalu_data-efficient_2024}. We randomly sampled 500 frames for each tripeptide and then trained the prior via score-matching until convergence. The parameters for the bond, angle, and dihedral terms were then fixed and the model was further trained on the octapeptide dataset from Ref.~\cite{lewis_scalable_2024} via score-matching until convergence. The resulting parameters were used to initialize the prior with the parameters for the bond, angle, and dihedral terms fixed when training the full CG model.

\subsubsection{Model Training}
All models used the AdamW optimizer with a learning rate of $10^{-3}$ and were trained on 2-4 NVIDIA H100 GPUs. Following CGSchNet \cite{charron_navigating_2025}, we employed the radial basis functions from PhysNet \cite{unke_physnet_2019} in our SchNet models. A The SchNet and MACE models were trained until convergence, with specific hyperparameters shown below. 

The eSEN architecture supports two readout options: a direct-force prediction that is trained separately from the energy prediction or a conservative force prediction computed as the negative gradient of the energy. While the direct-force model is non-conservative, it improves model efficiency. To take advantage of this during training, we followed the eSEN training protocol as described in Ref. \cite{fu_learning_2025}: models were first pre-trained for 60 epochs using direct-force prediction, after which the direct-force head was removed and the model was fine-tuned using conservative force prediction. Specific hyperparameters for the eSEN models are shown below. 

\begin{table}[h]
\centering
\begin{tabular}{c|c}
    \multicolumn{2}{c}{SchNet Model Hyperparameters} \\ \hline
    Feature embedding size & 128 \\[6pt]
    Number of filters & 128 \\[6pt]
    Interaction blocks & 2 \\[6pt]
    Radial basis & PhysNet \cite{unke_physnet_2019} \\[6pt]
    Number of basis functions & 64 \\[6pt]
    Cutoff & 20 \AA \\[12pt] 
    \multicolumn{2}{c}{MACE Model Hyperparameters} \\ \hline
    Feature Channels & 32 \\[6pt]
    Layers & 2 \\[6pt]
    $l_{max}$ & 3 \\[6pt]
    Number of radial basis functions & 8 \\[6pt]
    Readout hidden dimension & 16 \\[6pt]
    Cutoff & 20 \AA \\[12pt]
    \multicolumn{2}{c}{eSEN Model Hyperparameters} \\ \hline
    Sphere channels & 16 \\[6pt]
    Edge channels & 16 \\[6pt]
    Hidden channels & 16 \\[6pt]
    Layers & 2 \\[6pt]
    $l_{max}$ & 2 \\[6pt]
    Radial basis & Gaussian \\[6pt]
    Number of radial basis functions & 10 \\[6pt]
    Cutoff & 20 \AA \\[6pt]
\end{tabular} 
\end{table}

\begin{table}[h]
\centering
\begin{tabular}{c|cccc}
    \multicolumn{5}{c}{SchNet Training Hyperparameters} \\ \hline
    & MFM & MFM 100K & FM & SM \\ \hline
    Batch size & 96 & 96 & 96 & 48 \\[6pt]
    Number of GPUs & 2 & 2 & 4 & 4 \\[12pt] 
    \multicolumn{5}{c}{MACE Training Hyperparameters} \\ \hline
    & MFM & MFM 100K & FM & SM \\ \hline
    Batch size & 32 & 32 & 32 & 32 \\[6pt]
    Number of GPUs & 4 & 4 & 4 & 4 \\[12pt]
    \multicolumn{5}{c}{eSEN Training Hyperparameters} \\ \hline
    & MFM & MFM 100K & FM & SM \\ \hline
    Direct-force batch size & 64 & 64 & 64 & 48 \\[6pt]
    Conservative force batch size & 8 & 8 & 8 & 6 \\[6pt]
    Number of GPUs & 4 & 4 & 4 & 4 \\[6pt]
\end{tabular}
\end{table}

\begin{figure}
    \centering
    \includegraphics[width=0.5\linewidth]{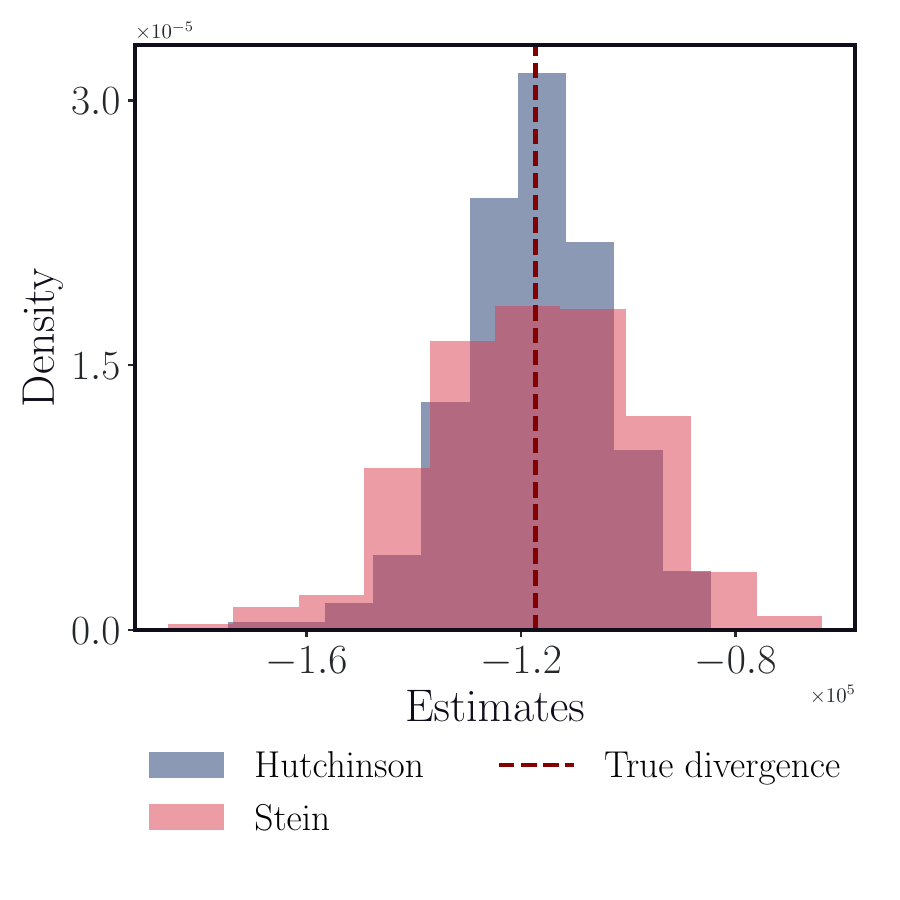}
    \caption{Histogram of estimated trace values using Hutchinson's trace estimator and Stein's Lemma. Histograms are constructed for 500 estimates using a randomly selected configuration and the MACE architecture. }
    \label{fig:trace}
\end{figure}

The score-matching objective requires computing the Laplacian $\Delta U_{\theta}$, which is computationally expensive. To reduce this cost, we use a form of Hutchinson's trace estimator, referred to as Stein's trace estimator, as described in Eq. \ref{eq:trace}. Fig. \ref{fig:trace} compares the variance between the Hutchinson and Stein estimators by plotting the distribution of 500 trace estimates. Although Stein's trace estimator exhibits larger variance, it was used for training all SM models due to its improved efficiency. All SM models used $N_{\textrm{sample}} = 1$ and $\epsilon = 0.0001$.

\subsection{Atomistic Free Energy Calculations}
\subsubsection{Trp-cage}
The native contact collective variable for Trp-cage (PDB: 1L2Y) was computed by first identifying $\textrm{C}_{\alpha}$ pairs that were separated by 2 or more residues in sequence with a pairwise distance less than 8 \AA~in the native structure. To ensure smoothness, the native contact collective variable was then computed as
\begin{equation*}
    \frac{1}{N} \sum_{i,j}^{\textrm{C}_{\alpha}\textrm{ pairs}} \frac{1}{1 + e^{-2 (5.25 - ||\boldsymbol{r_i} - \boldsymbol{r_j}||^2_2)}}
\end{equation*}
where $N$ is the total number of native contact pairs, $\boldsymbol{r}_i$ and $\boldsymbol{r}_j$ are the positions of $\textrm{C}_{\alpha}$ atoms $i$ and $j$, and the switching function transitions smoothly from 1 (contact formed) to 0 (contact broken) around a cutoff distance of 5.25 \AA.

Four simulations were performed with starting configurations selected from Ref. \cite{lindorff-larsen_how_2011}, one starting in each of the four metastable basins shown in Fig. \ref{fig:trpcage}. Each structure was solvated using the TIP3P water model, with a 10 \AA~cubic buffer. Simulations were performed using the AMBER ff99SBxildn force field and OpenMM (version 8.1.1), with the OVRVO integrator at 300 K, a 4 fs time step, and a friction coefficient of 1.0 $\textrm{ps}^{-1}$. Frames were saved every 20 ps and each simulation was run for 13-16 $\mu \textrm{s}$, yielding over 60 $\mu \textrm{s}$ of aggregate simulation.

\subsubsection{BBA}
Using starting configurations from BBA (PDB 1FME) trajectories reported in Ref. \cite{majewski_machine_2023}, we ran over 170 $\mu\textrm{s}$ of unbiased simulation in explicit solvent using the AMBER ff99SBdisp forcefield and OpenMM. For more rigorous free energy calculation, umbrella sampling simulations with 100 windows were performed using the PLUMED plugin for OpenMM. Umbrella centers were chosen at evenly spaced intervals across the FES produced by the unbiased MD as shown in Fig. \ref{fig:umbrella}. Starting configurations for each window were chosen as close as possible to the corresponding center.
Each structure was solvated using the TIP3P water model, with a 9 \AA cubic buffer at 150 mM salt concentration. A harmonic restraint with a force constant of 50 kcal/mol was applied to maintain sampling near each window center. Each umbrella window underwent a 50 ns equilibration period followed by 450 ns of production sampling, with frames saved every 1.5 ns. The resulting trajectories were analyzed using PyMBAR to compute free energy surface via the Multistate Bennett Acceptance Ratio

\subsection{CG Simulations}
\begin{figure}
    \centering
    \includegraphics[width=0.49\linewidth]{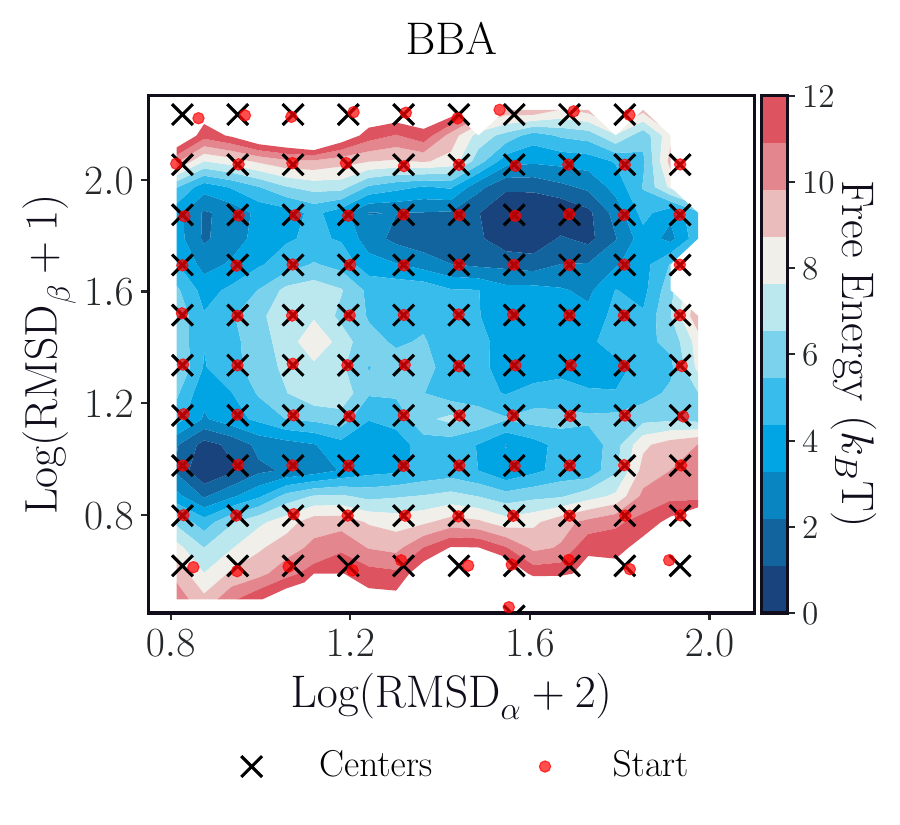}
    \includegraphics[width=0.49\linewidth]{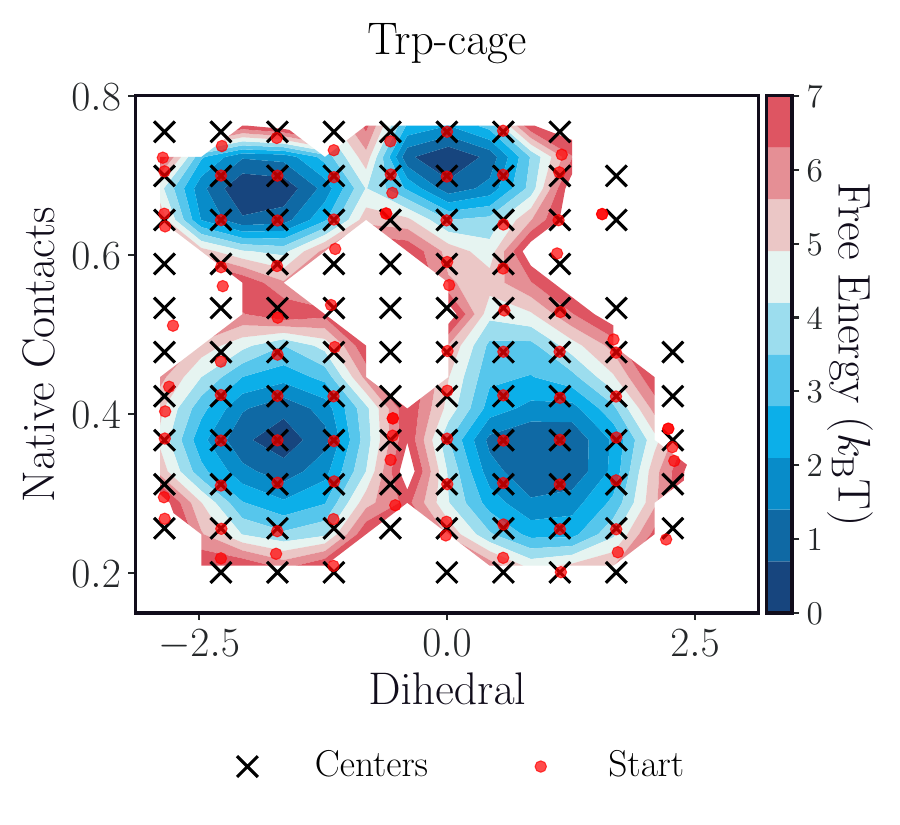} 
    \caption{ Umbrella sampling centers in black and starting configurations in red for (a) atomistic BBA FES produced via umbrella sampling and (b) Atomistic Trp-cage FES. }
    \label{fig:umbrella}
\end{figure}

To produce the CG FES for Trp-cage, umbrella sampling simulations with 100 windows were performed.  As shown in Fig. \ref{fig:umbrella}, umbrella centers were selected at evenly spaced intervals across the atomistic FES. Starting configurations for each window were chosen as close as possible to the corresponding center. A harmonic restraint with a force constant of 25 kcal/mol was applied to maintain sampling near each window center. Simulations were performed with a time step of 0.2 ps and friction coefficient of 1 $\textrm{ps}^{-1}$. Each umbrella window underwent a 20 ns equilibration period followed by 180 ns of production sampling, with frames saved every 400 ps. The resulting trajectories were analyzed using PyMBAR to compute free energy surface via the Multistate Bennett Acceptance Ratio.

An identical procedure was performed to produce the CG FES for BBA. The same window centers and starting configurations used for the atomistic umbrella sampling for BBA shown in Fig. \ref{fig:umbrella} were used for the CG umbrella sampling.

\subsection{Comparison to other transferable CG}
\begin{figure}
    \centering
    \includegraphics[width=0.8 \linewidth]{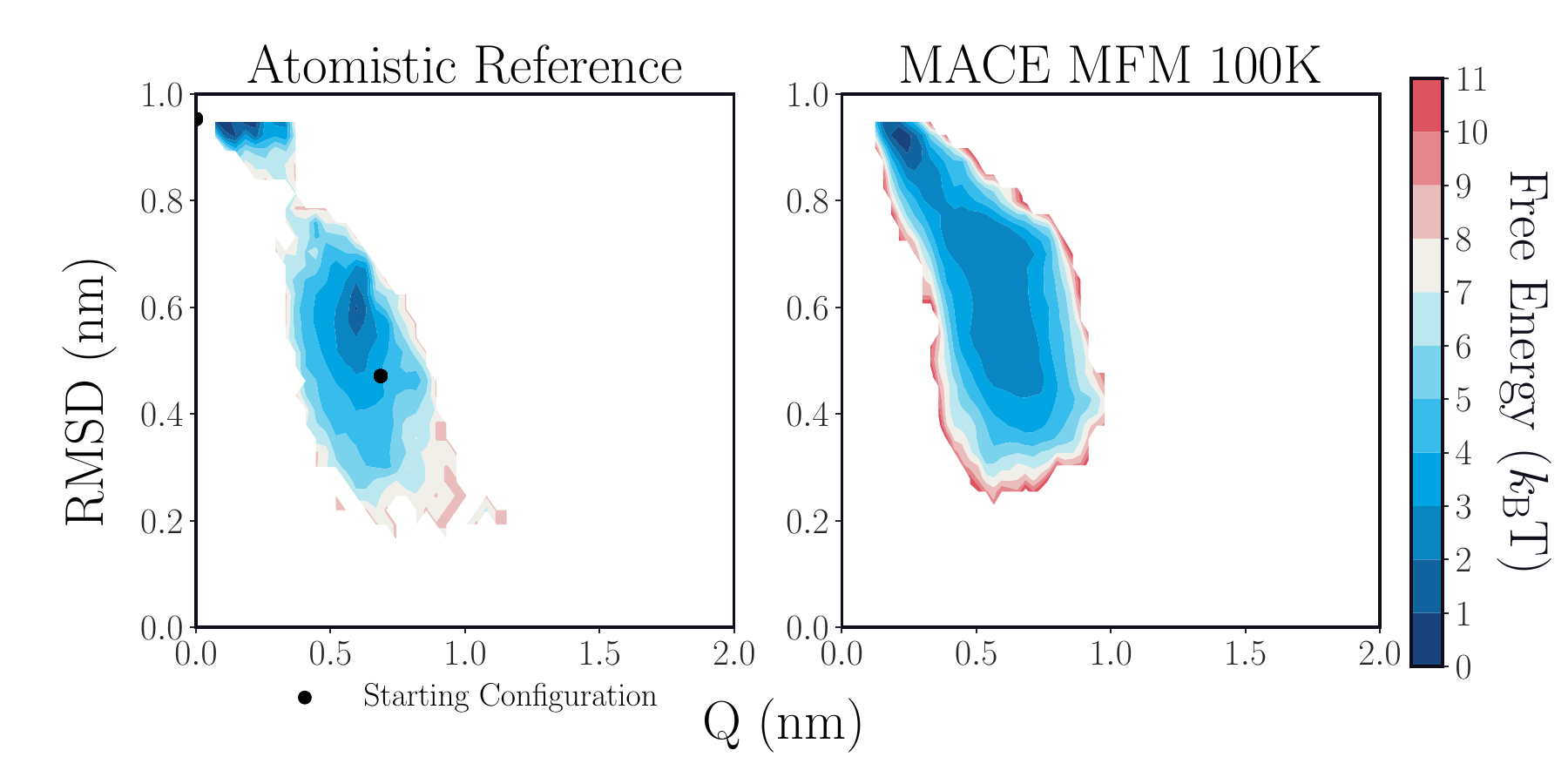}
    \caption{FES via Langevin Simulations of Trp-cage.}
    \label{fig:comp_trp}
\end{figure}

\begin{figure}
    \centering
    \includegraphics[width=0.8 \linewidth]{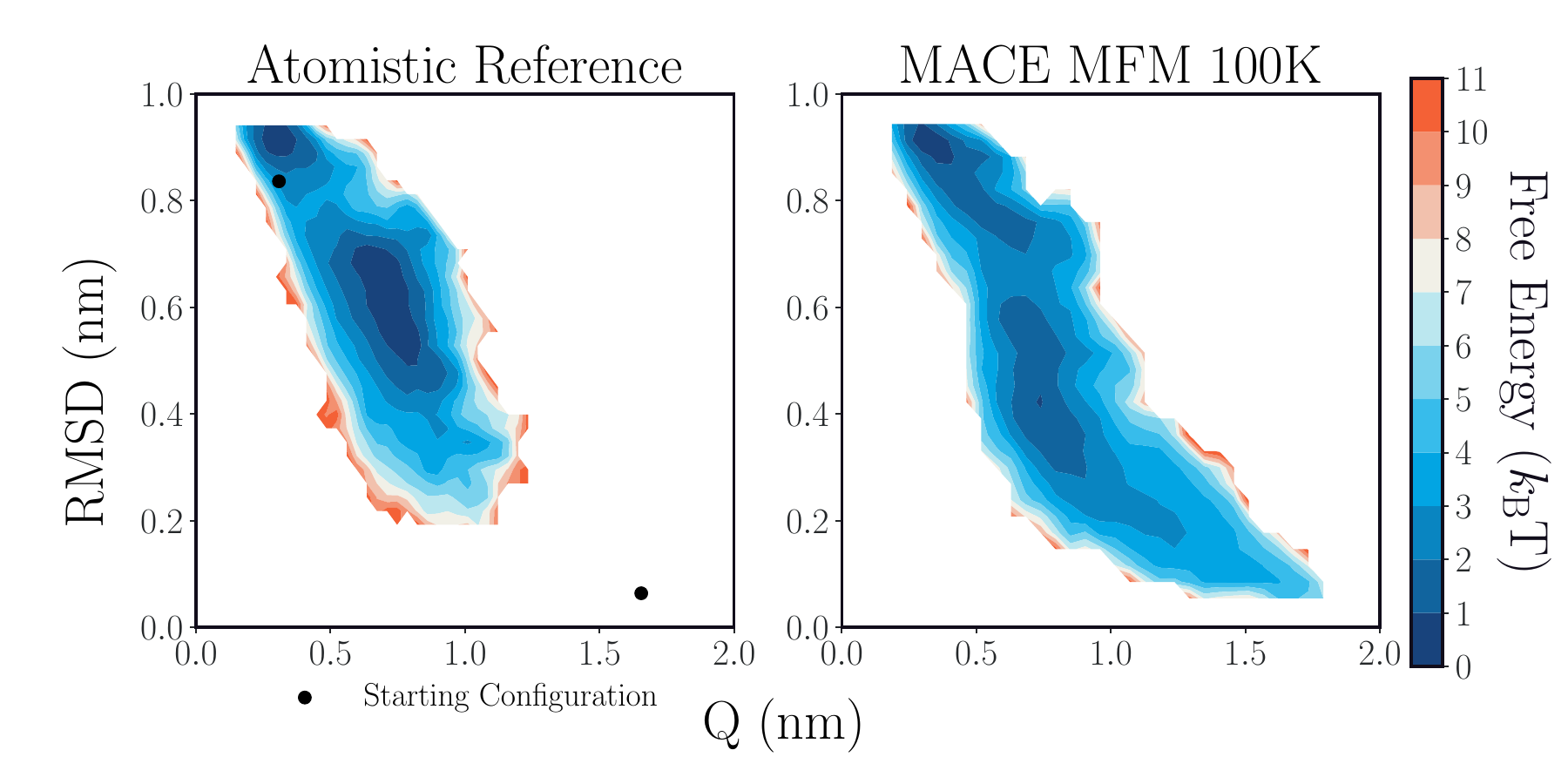}
    \caption{FES via Langevin Simulations of BBA.}
    \label{fig:comp_bba}
\end{figure}

We sought to directly compare our CG models to the transferable CG model from Charron et al. \cite{charron_navigating_2025}. Following their procedure for Langevin simulations for Trp-cage, 35 simulations were started in a folded configuration and 35 simulations were started in an unfolded configuration. For BBA, 25 simulations were started in a folded configuration and 25 simulations in an unfolded configuration. Langevin simulations were run for 4 million steps with a 100,000 step burn-in.  To enable direct comparison with the CG FES reported in Supplementary Fig. 19 of Ref. \cite{charron_navigating_2025}, we used their reported native contact and RMSD collective variables to construct the free energy surfaces shown in Fig. \ref{fig:comp_trp} and Fig. \ref{fig:comp_bba}. Our MACE MFM-100K model shows good agreement with the atomistic reference.

\end{document}